\documentclass[a4paper,11pt]{article}
\usepackage{jcappub}
\usepackage{lineno}
\usepackage{float}
\usepackage[english]{babel}
\usepackage{subfloat}
\usepackage{subcaption}
\usepackage{amsmath}
\usepackage{amssymb}
\usepackage{graphicx}
 \usepackage{multirow}
\title{Search for non-standard neutrino interactions with the first six detection units of KM3NeT/ORCA}

\author[a]{S.~Aiello}
\author[b,bb]{A.~Albert}
\author[c]{A.\,R.~Alhebsi}
\author[d]{M.~Alshamsi}
\author[e]{S. Alves Garre}
\author[g,f]{A. Ambrosone}
\author[h]{F.~Ameli}
\author[i]{M.~Andre}
\author[j]{L.~Aphecetche}
\author[k]{M. Ardid}
% ...
\emailAdd{Alfonso.Lazo@ific.uv.es}
\emailAdd{km3net-pc@km3net.de}
\author[d]{S.~Le~Stum}
\author[o]{G.~Lehaut}
\author[ak]{V.~Lema{\^\i}tre}
\author[a]{E.~Leonora}
\author[e]{N.~Lessing}
\author[s,r]{G.~Levi}
\author[l]{M.~Lindsey~Clark}
\author[a]{F.~Longhitano}
\author[d]{F.~Magnani}
\author[w]{J.~Majumdar}
% ...

\affiliation[a]{INFN, Sezione di Catania, (INFN-CT) Via Santa Sofia 64, Catania, 95123 Italy}
\affiliation[b]{Universit{\'e}~de~Strasbourg,~CNRS,~IPHC~UMR~7178,~F-67000~Strasbourg,~France}
\affiliation[c]{Khalifa University, Department of Physics, PO Box 127788, Abu Dhabi, 0000 United Arab Emirates}
\affiliation[d]{Aix~Marseille~Univ,~CNRS/IN2P3,~CPPM,~Marseille,~France}
\affiliation[e]{IFIC - Instituto de F{\'\i}sica Corpuscular (CSIC - Universitat de Val{\`e}ncia), c/Catedr{\'a}tico Jos{\'e} Beltr{\'a}n, 2, 46980 Paterna, Valencia, Spain}
\affiliation[f]{INFN, Sezione di Napoli, Complesso Universitario di Monte S. Angelo, Via Cintia ed. G, Napoli, 80126 Italy}
\affiliation[g]{Universit{\`a} di Napoli ``Federico II'', Dip. Scienze Fisiche ``E. Pancini'', Complesso Universitario di Monte S. Angelo, Via Cintia ed. G, Napoli, 80126 Italy}
\affiliation[h]{INFN, Sezione di Roma, Piazzale Aldo Moro 2, Roma, 00185 Italy}
\affiliation[i]{Universitat Polit{\`e}cnica de Catalunya, Laboratori d'Aplicacions Bioac{\'u}stiques, Centre Tecnol{\`o}gic de Vilanova i la Geltr{\'u}, Avda. Rambla Exposici{\'o}, s/n, Vilanova i la Geltr{\'u}, 08800 Spain}
\affiliation[j]{Subatech, IMT Atlantique, IN2P3-CNRS, Nantes Universit{\'e}, 4 rue Alfred Kastler - La Chantrerie, Nantes, BP 20722 44307 France}
\affiliation[k]{Universitat Polit{\`e}cnica de Val{\`e}ncia, Instituto de Investigaci{\'o}n para la Gesti{\'o}n Integrada de las Zonas Costeras, C/ Paranimf, 1, Gandia, 46730 Spain}
\affiliation[l]{Universit{\'e} Paris Cit{\'e}, CNRS, Astroparticule et Cosmologie, F-75013 Paris, France}
% ...
\affiliation[o]{LPC CAEN, Normandie Univ, ENSICAEN, UNICAEN, CNRS/IN2P3, 6 boulevard Mar{\'e}chal Juin, Caen, 14050 France}
% ...
\affiliation[r]{INFN, Sezione di Bologna, v.le C. Berti-Pichat, 6/2, Bologna, 40127 Italy}
\affiliation[s]{Universit{\`a} di Bologna, Dipartimento di Fisica e Astronomia, v.le C. Berti-Pichat, 6/2, Bologna, 40127 Italy}
% ...
\affiliation[w]{Nikhef, National Institute for Subatomic Physics, PO Box 41882, Amsterdam, 1009 DB Netherlands}
% ...
\affiliation[ak]{UCLouvain, Centre for Cosmology, Particle Physics and Phenomenology, Chemin du Cyclotron, 2, Louvain-la-Neuve, 1348 Belgium}
% ...
\affiliation[bb]{Universit{\'e} de Haute Alsace, rue des Fr{\`e}res Lumi{\`e}re, 68093 Mulhouse Cedex, France}
% ...

\abstract{KM3NeT/ORCA is an underwater neutrino telescope under construction in the Mediterranean Sea. Its primary scientific goal is to measure the atmospheric neutrino oscillation parameters and to determine the neutrino mass ordering. ORCA can constrain the oscillation parameters $\Delta m^{2}_{31}$ and $\theta_{23}$ by reconstructing the arrival direction and energy of multi-GeV neutrinos crossing the Earth. Searches for deviations from the Standard Model of particle physics in the forward scattering of neutrinos inside Earth matter, produced by Non-Standard Interactions, can be conducted by investigating distortions of the standard oscillation pattern of neutrinos of all flavours. This work reports on the results of the search for non-standard neutrino interactions using the first six detection units of ORCA and 433 kton-years of exposure. No significant deviation from standard interactions was found in a sample of 5828 events reconstructed in the 1 GeV$-$1 TeV energy range. The flavour structure of the non-standard coupling was constrained at 90\% confidence level to be $|\varepsilon_{\mu\tau} | \leq 5.4 \times 10^{-3}$, $|\varepsilon_{e\tau} | \leq 7.4 \times 10^{-2}$, $|\varepsilon_{e\mu} | \leq 5.6 \times 10^{-2}$ and $-0.015 \leq \varepsilon_{\tau\tau} - \varepsilon_{\mu\mu} \leq 0.017$. The results are comparable to the current most stringent limits placed on the parameters by other experiments.}

\begin{document}
\nocite{*}
\maketitle
\flushbottom

%\input{introduction.tex}
%%%%%%%%%%%%%%%%%%%%%%%%%%%%%%%%%%%%%%%%%%%%%%%%%%%%%%%%%%%%%%%%%%%%%%%%%%%%%%%%%%%%%%%%%%%%%%%
\section{Introduction}
\label{sec:introduction}

Neutrino propagation is described as the evolution of a coherent superposition of the different mass eigenstates of the Hamiltonian. The mixing makes the transition probability between the produced and detected neutrino flavour eigenstates exhibit an oscillatory dependence, in the phenomenon referred to as neutrino oscillations. The first observations of solar \cite{SNO:2001kpb} and atmospheric neutrino oscillations \cite{Super-Kamiokande:1998kpq, MACRO:1998ckv} implied that at least two neutrino mass eigenstates carry non-zero mass, in direct contradiction with the assumption of the electroweak theory. The neutrino mass-squared differences have been measured with increasing precision over the years \cite{Denton:2022een}, and current global analyses of neutrino oscillation data constrain them to the few-percent level \cite{ParticleDataGroup:2024cfk, Esteban:2020cvm, deSalas:2020pgw, Capozzi:2021fjo}. However, the underlying mechanism giving rise to neutrino masses remains unknown.

Most models proposed to account for the origin of neutrino masses imply the existence of new-physics mediators, acting above some characteristic energy scale $\Lambda$. Their effects are formally introduced by considering the Standard Model (SM) as part of an Effective Field Theory (EFT), such that after integrating the heavy degrees of freedom above $ \Lambda$, the model would only invoke the fields relevant at the scale of the SM, respecting its symmetries \cite{Langacker:2010zza}. Within such a framework, the effective Lagrangian can be parameterised in terms of non-renormalisable operators as in \cite{Bischer:2019ttk}:

\begin{equation}\label{eq:expansion}
    \mathcal{L}_{\text {eff }}=\mathcal{L}_{\mathrm{SM}}+\sum_i \sum_{n \geq 5} \frac{1}{\Lambda^{n-4}} C_i \mathcal{O}_i^{(n)} ,
\end{equation}

\noindent where $i$ runs over all operators $\mathcal{O}_i^{(n)}$ at order $n\geq 5$ allowed by the lower-energy symmetries, and $C_i$ are dimensionless coefficients. The order $n=5$ is the Weinberg operator responsible for neutrino masses after electroweak symmery breaking is realised \cite{Weinberg:1979sa, Grzadkowski:2010es}, while $n\geq6$ involve new forms of interactions denoted as Non-Standard Interactions (NSI). Orders higher than $n=8$ are heavily suppressed by decreasing factors $1/\Lambda^{n-4}$, thus are not expected to produce sizeable effects below $\Lambda$. 

The higher-order operators can lead to both charged-current (CC) and neutral-current (NC) neutrino ineractions. The NSI relevant for atmospheric neutrino propagation affect the NC forward scattering of neutrinos on fermions found in ordinary matter. Such interactions would arise from allowed Lagrangian terms involving four fermions, out of which two are required to be left-handed neutrino fields \cite{Bischer:2019ttk}, restricting the most general form of the NC NSI term relevant in neutrino propagation in a model-independent way to \cite{Farzan:2017xzy}

\begin{equation}\label{eq:lagrangian}
    \mathcal{L}_{\mathrm{NSI}}^{\mathrm{NC}}=-2 \sqrt{2} G_F \varepsilon_{\alpha \beta}^{f X}\left(\bar{\nu}_\alpha \gamma_\mu P_L \nu_\beta\right)\left(\bar{f} \gamma^\mu P_X f\right) ,
\end{equation}

\noindent where $G_F$ is the Fermi constant, $X$ runs over the left- and right-chirality projection operators $P_{L,R}=(1 \pm \gamma_5)/2$, $\alpha $ and $\beta$ run over the neutrino flavours $ \{e,\mu,\tau\}$ and $f$ runs over the fermions found in ordinary matter $\{e,u ,d\}$. The dimensionless complex parameters $\varepsilon_{\alpha \beta}$ are the NSI coupling strengths involved in $\nu_\alpha \rightarrow \nu_\beta$ transitions, defined relative to the standard electroweak magnitude given by $G_F$. Standard neutrino interactions are recovered in the limit $\varepsilon_{\alpha \beta}\rightarrow 0$, whereas $\varepsilon_{\alpha \beta}\sim 1$ would be indicative of interaction strengths comparable to those involving SM processes. The presence of non-zero NSI couplings therefore enriches the SM phenomenology by enabling processes not allowed within its scope. In particular,  $\varepsilon_{\alpha\alpha}\neq \varepsilon_{\beta \beta} $ implies non-universal lepton couplings in the NC interaction of neutrinos, while $\varepsilon_{\alpha\beta}\neq 0$ for $\alpha \neq \beta$ lead to flavour-changing neutral currents. 

For the neutrino propagation through a static, unpolarised medium such as neutral Earth matter, the chirality sum over $X \in \{L,R\}$ in the fermion vertex of eq.~\eqref{eq:lagrangian} is averaged out, retaining only vector-like NSI couplings with corresponding strengths $\varepsilon_{\alpha\beta}^{fV} = \varepsilon_{\alpha\beta}^{fL} + \varepsilon_{\alpha\beta}^{fR}$. Moreover, neutrinos traversing Earth matter scatter off electrons, up and down quarks during their NC forward scattering, resulting in neutrino propagation sensitive to the incoherent sum of the scattering amplitudes on the three types of fermions \cite{Chatterjee:2014gxa}:

\begin{equation}\label{eq:NSIfermion}
    \varepsilon_{\alpha \beta} = \varepsilon_{\alpha \beta}^{eV}+ Y_{u} \varepsilon_{\alpha \beta}^{u V}+Y_{d} \varepsilon_{\alpha \beta}^{d V} ,
\end{equation}

\noindent with $Y_u = N_u/N_e$ and $Y_d = N_d/N_e$ the fraction of the NSI strength attributed to each target fermion, based on their number density relative to electrons. Following the Preliminary Reference Earth Model (PREM) \cite{Dziewonski:1981xy} which fixes the averaged Earth composition, one obtains $Y_u \approx 3.051$ and $Y_d \approx 3.102$~\cite{Gonzalez-Garcia:2013usa}. In the following, down quarks are assumed to fully account for NSI couplings, neglecting electron and up-quark contributions. Comparisons to other experiments will then be performed in terms of the effective NSI couplings from eq.~\eqref{eq:NSIfermion} by applying the corresponding scaling factors. 

This paper presents the results from the search for NSI with 433 kton-years of exposure of the ORCA detector with its first six Detection Units (DUs), called hereafter ORCA6. ORCA is a water Cherenkov neutrino telescope under construction in the Mediterranean Sea, with the main goal of a precision measurement of the atmospheric neutrino oscillation parameters and the determination of the neutrino mass ordering (NMO). 

This paper is structured as follows: in section \ref{sec:NSIprobs} the phenomenology behind neutrino oscillations in the presence of NSI is explained. Then, the detector configuration and the event sample are presented in sections \ref{sec:detector} and \ref{sec:dselection}, respectively. The statistical approach followed in the analysis is covered in section \ref{sec:analysis}, and the results are shown in section \ref{sec:results}. Conclusions are then drawn in section \ref{sec:conclusions}.
% FOR THE ABSTRACT.
% ORCA will host 115 DUs upon completion \cite{KM3Net:2016zxf}, and is expected to detect $\sim6\times 10^5$ events per year after quality and selection criteria \cite{KM3NeT:2021ozk}. By inspecting neutrino arrival directions from horizontal to upgoing-vertical trajectories, ORCA can effectively constrain NSI parameters based on the neutrino oscillation pattern, as they undergo NC forward scattering while traversing different Earth density profiles.

%%%%%%%%%%%%%%%%%%%%%%%%%%%%%%%%%%%%%%%%%%%%%%%%%%%%%%%%%%%%%%%%%%%%%%%%%%%%%%%%%%%%%%%%%%%%%%%%%%%%%%%%%%%%%%%%%%%

%\input{NSIprobs.tex}
\section{Neutrino oscillation probabilities in the presence of NSI}
\label{sec:NSIprobs}

Neutrino flavour oscillations in vacuum are governed by the squared-mass splittings $\Delta m^{2}_{ij}$ and mixing angles $\theta_{ij}$, where $i,j\in \{1,2,3\}$ denote the mass eigenstates of the propagation Hamiltonian. In the presence of matter, the coherent forward scattering of neutrinos off the background fermions exhibits a net flavour asymmetry, as electron neutrinos undergo additional CC forward scattering not taking place for muon and tau neutrinos, as pointed out by Wolfenstein \cite{Wolfenstein:1977ue}. Within the SM framework, the $\nu_e/\bar{\nu}_e$ CC interaction with the target fermion integrates into the flavour-diagonal, position-dependent matter potential \allowbreak $2\sqrt{2}G_F N_e (x)\times \mathrm{diag}\{1,0,0\}$~\cite{RevModPhys.61.937}, with $N_e (x)$ the electron number density along the neutrino path. Under NSI, the forward NC scattering of neutrinos could be flavour-violating and non-universal among neutrino flavours, consequently leading to a modified effective neutrino propagation Hamiltonian in matter, which in the ultrarelativistic limit reads \cite{Wang:2018dwk,Coloma:2023ixt,Esteban:2018ppq,Proceedings:2019qno,Coloma:2016gei,Friedland:2004ah,Feng:2019mno,Kikuchi:2008vq,Esmaili:2013fva}:

\begin{equation}\label{eq:NSIham}
    H_{\mathrm{eff}}=\frac{1}{2 E} \ \mathcal{U}\left[\begin{array}{ccc}
0 & 0 & 0 \\
0 & \Delta m_{21}^2 & 0 \\
0 & 0 & \Delta m_{31}^2
\end{array}\right] \mathcal{U}^{+}+A(x)\left[\begin{array}{ccc}
1+\varepsilon_{e e} & \varepsilon_{e \mu} & \varepsilon_{e \tau} \\
\varepsilon_{e \mu}^* & \varepsilon_{\mu \mu} & \varepsilon_{\mu \tau} \\
\varepsilon_{e \tau}^* & \varepsilon_{\mu \tau}^* & \varepsilon_{\tau \tau}
\end{array}\right] \text {, }
\end{equation}

\noindent where $E$ is the neutrino energy, $\mathcal{U}$ is the PMNS matrix encoding the parameterisation of the neutrino mixing \cite{Maki:1962mu}, $A(x)=\sqrt{2}G_F N_e (x)$ is the matter potential and $\varepsilon_{\alpha \beta}=|\varepsilon_{\alpha \beta}|e^{i\delta_{\alpha\beta}}$ are the effective NSI coupling strengths from eq.~\eqref{eq:NSIfermion}. For antineutrinos, eq.~\eqref{eq:NSIham} transforms under $\mathcal{U} \rightarrow \mathcal{U}^*$, $A(x) \rightarrow -A(x)$ and $\varepsilon_{\alpha \beta} \rightarrow \varepsilon^{*}_{\alpha \beta}$ \cite{Wang:2018dwk}. 

For vacuum propagation, $A(x)=0$ and eq.~\eqref{eq:NSIham} exhibits the degeneracies

\begin{equation}\label{eq:vacuumtrans}
    \begin{array}{l}
\Delta m_{31}^2 \leftrightarrow-\Delta m_{31}^2+\Delta m_{21}^2=-\Delta m_{32}^2 \text , \\
 \theta_{12} \leftrightarrow \pi/2 - \theta_{12} \text , \\
\delta_{\mathrm{CP}} \leftrightarrow \pi-\delta_{\mathrm{CP}} \text ,
\end{array}
\end{equation}

\noindent which leave the flavour transition probabilities invariant. The presence of $A(x)$ in matter propagation breaks the above-mentioned degeneracy assuming standard interactions (SI), i.e. $\varepsilon_{\alpha\beta} = 0 $.  However, the full generalised matter potential under NSI from eq.~\eqref{eq:NSIham} recovers the vacuum degeneracy broken by the standard term alone. The transformations eq.~\eqref{eq:vacuumtrans} together with the mapping of the NSI parameters 

\begin{equation}\label{eq:nsitrans}
    \begin{array}{l}
\left(\varepsilon_{e e}-\varepsilon_{\mu \mu}\right) \rightarrow-\left(\varepsilon_{e e}-\varepsilon_{\mu \mu}\right)-2 \text , \\
\left(\varepsilon_{\tau \tau}-\varepsilon_{\mu \mu}\right) \rightarrow-\left(\varepsilon_{\tau \tau}-\varepsilon_{\mu \mu}\right) \text , \\
\epsilon_{\alpha \beta} \rightarrow-\epsilon_{\alpha \beta}^* \quad(\alpha \neq \beta) \text ,
\end{array}
\end{equation}

\noindent map $H_{\mathrm{eff}} \rightarrow -H^{*}_{\mathrm{eff}}$, therefore leave the neutrino evolution equation in matter invariant under complex conjugation of the amplitudes \cite{Coloma:2016gei}. The oscillation probabilities depend on differences in the diagonal elements of the NSI Hamiltonian (see eqs.~\eqref{eq:NSImumu} and \eqref{eq:NSImue}), for which reason only $\varepsilon_{\tau\tau}-\varepsilon_{\mu\mu}$ is observable in the analysis, and diagonal elements of eq. \ref{eq:NSIham} are real valued. The degeneracy with the NMO as seen in eq.~\eqref{eq:vacuumtrans} is however not exact in the present analysis, given the position dependence of fermion densities inside the Earth (see eq. \eqref{eq:NSIfermion}) \cite{Esteban:2018ppq} and that  NSI couplings are considered one by one in the fits to the data. Moreover, the difference $\varepsilon_{e e}-\varepsilon_{\mu \mu}$ is fixed at 0 in the analysis, since negligible sensitivity is expected from MC studies.  %% HMMM REPHRASE BETTER??

\begin{figure}[t!]
\centering
\includegraphics[width=0.90\textwidth]{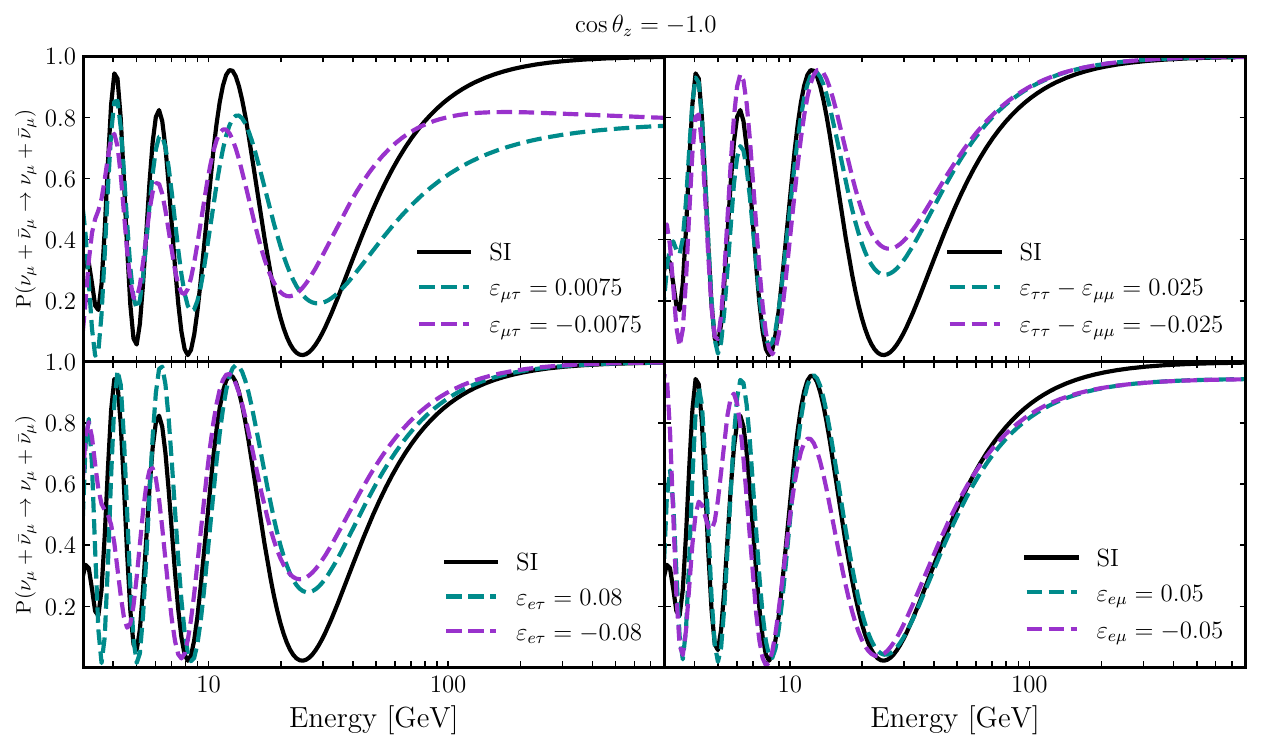}
\qquad
\caption{Combined $\nu_{\mu}+\bar{\nu}_\mu$ survival probability for neutrinos propagating along the up-going, Earth core-crossing trajectory $\cos \theta_z=-1$. The standard interaction prediction (SI) is represented by the solid black line, and is identical for the four subplots. Dashed lines show the NSI expectation for positive and negative coupling strength in the cases of non-zero $\varepsilon_{\mu\tau}$ (top left), $\varepsilon_{\tau\tau} - \varepsilon_{\mu\mu}$ (top right), $\varepsilon_{e\tau}$ (bottom left) and $\varepsilon_{e\mu}$ (bottom right). The specific values of NSI parameters were chosen for visualisation purposes. For all curves, the standard oscillation parameters $\Delta m^{2}_{ij}$ and $\theta_{ij}$ were fixed at NuFit 5.0 \cite{Esteban:2020cvm} for normal ordering (NO). }
\label{fig:NuMuProbs}
\end{figure}

\begin{figure}[t!]
\centering
\includegraphics[width=0.9\textwidth]{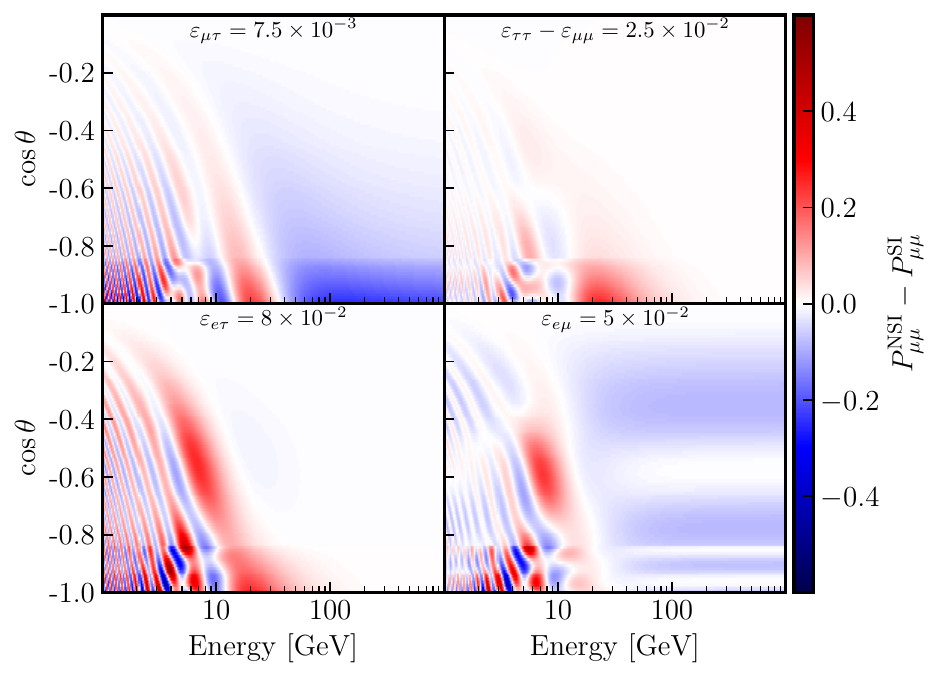}
\qquad
\caption{ Difference in combined $\nu_\mu + \bar{\nu}_\mu$ survival probability between four different NSI realisations, $P^{\mathrm{NSI}}_{\mu\mu}$, and the expectation from standard interactions $P^{\mathrm{SI}}_{\mu\mu}$, as a function of the neutrino energy and arrival direction. NuFit 5.0 \cite{Esteban:2020cvm} at NO is assumed for the standard oscillation parameters. Blue-shaded areas would indicate a decrease in $\nu_\mu$ and $\bar{\nu}_\mu$ events when NSI is present in comparison to SI; an event excess would apply to red-shaded areas. From left to right, the NSI realisations assume real-valued $\varepsilon_{\mu\tau}$, $\varepsilon_{\tau\tau} - \varepsilon_{\mu\mu}$, $\varepsilon_{e\tau}$ and $\varepsilon_{e\mu}$ while fixing the rest to zero. }
\label{fig:NuMuOsc}
\end{figure}

Sections \ref{sec:NSImutau} and \ref{sec:NSIelec} cover a discussion of the phenomenology arising in the presence of NSI, keeping only one coupling strength to be non-zero at a time for simplification. The exact flavour transition probabilities can be obtained by numerical diagonalisation of the Hamiltonian from eq.~\eqref{eq:NSIham} using the OscProb software package \cite{joao_coelho_2023_8074017}. In Figs.~\ref{fig:NuMuProbs}, \ref{fig:NuEProbs} and \ref{fig:NuMuOsc} the software is employed to compute the combined transition amplitudes of muon neutrinos and antineutrinos, re-weighting each chirality as $P^{\nu_{\mu}+\bar{\nu}_\mu \rightarrow \nu_{x}+\bar{\nu}_x} = ( P^{\nu_{\mu}\rightarrow \nu_{x}}+0.5\times P^{\bar{\nu}_{\mu}\rightarrow \bar{\nu}_{x}})/1.5$, accounting for the approximate inclusive CC-interaction cross-section asymmetry $\sim$ 2:1 between $\nu$ and $\bar{\nu}$ \cite{ParticleDataGroup:2024cfk} in the energy range probed by ORCA6. This visualisation incorporates the washout of the NSI effects due to the mixture of opposite chiralities $\nu$ and $\bar{\nu}$ in the atmospheric flux, and the mentioned approximation is exclusively used for Figs.~\ref{fig:NuMuProbs}, \ref{fig:NuEProbs} and \ref{fig:NuMuOsc} and not at analysis level. The zenith angle $\cos \theta_z = -1.0$ corresponding to the largest neutrino baseline across the Earth ($L\sim 12700$ km) was chosen for Figs.~\ref{fig:NuMuProbs} and \ref{fig:NuEProbs}, in order to enhance significant NSI effects. 

The three-flavour transition amplitudes between the eigenstates of the Hamiltonian in eq.~\eqref{eq:NSIham} can be analytically derived under certain assumptions. This approximation will be helpful to support the underlying phenomenology of the ORCA6 results, although the analysis employs the numerical diagonalisation of eq.~\eqref{eq:NSIham} and does not make use of the explicit oscillation probabilities shown later. In \cite{Chatterjee:2014gxa}, the following analytical expressions are proposed for $ \Delta P = P^{\mathrm{SI}}-P^{\mathrm{NSI}}$, the difference in oscillation probability between standard and non-standard interactions, up to first order corrections in $\Delta m^{2}_{21}/\Delta m^{2}_{31}$:

\begin{figure}[t!]
\centering
\includegraphics[width=0.90\textwidth]{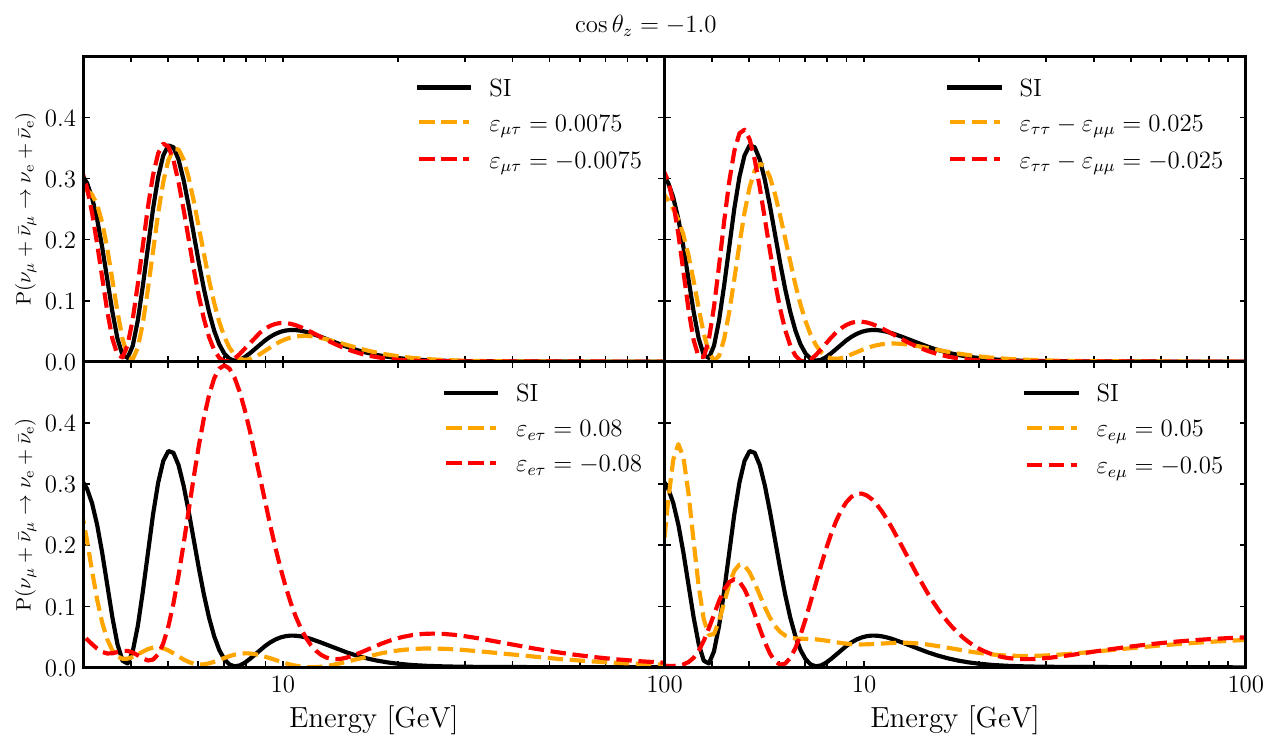}
\qquad
\caption{Combined $\nu_{\mu}+\bar{\nu}_\mu$ transition probability to $\nu_{e}+\bar{\nu}_e$, for neutrinos propagating along the up-going, Earth core-crossing trajectory $\cos \theta_z=-1$. The SI prediction is represented by the solid black line, and is identical for the four subplots. Dashed lines show the NSI expectation for positive and negative coupling strength in the cases of non-zero $\varepsilon_{\mu\tau}$ (top left), $\varepsilon_{\tau\tau} - \varepsilon_{\mu\mu}$ (top right), $\varepsilon_{e\tau}$ (bottom left) and $\varepsilon_{e\mu}$ (bottom right). The specific values of NSI parameters were chosen for visualisation purposes. For all curves, the standard oscillation parameters $\Delta m^{2}_{ij}$ and $\theta_{ij}$ were fixed at NuFit 5.0 \cite{Esteban:2020cvm} assuming NO. }
\label{fig:NuEProbs}
\end{figure}

\begin{equation}\label{eq:NSImumu}
    \begin{aligned}
\Delta P_{\mu \mu}= & | \varepsilon_{\mu \tau} | \cos \delta_{\mu \tau} s_{2 \times 23} \ r_A\left[s_{2 \times 23}^2\left(\lambda L\right) \sin \lambda L+4 c_{2 \times 23}^2 \sin ^2 \frac{\lambda L}{2}\right] \\
& -\left(\varepsilon_{\mu \mu}- \varepsilon_{\tau\tau}\right) s_{2 \times 23}^2 c_{2 \times 23} \ r_A\left[\frac{\lambda L}{2} \sin \lambda L -2 \sin ^2 \frac{\lambda L}{2}\right]
\end{aligned}
\end{equation}

\begin{equation}\label{eq:NSImue}
    \begin{aligned}
\Delta P_{e \mu}= & -8 s_{13} s_{23} c_{23}\left(\left|\varepsilon_{e \mu}\right| c_{23} c_X-\left|\varepsilon_{e \tau}\right| s_{23} c_\omega\right) r_A\left[\frac{\sin r_A \lambda L / 2}{r_A} \frac{\sin \left(1-r_A\right) \lambda L / 2}{\left(1-r_A\right)} \cos \frac{\lambda L}{2}\right] \\
& -8 s_{13} s_{23} c_{23}\left(\left|\varepsilon_{e \mu}\right| c_{23} s_X-\left|\varepsilon_{e \tau}\right| s_{23} s_\omega\right) r_A\left[\frac{\sin r_A \lambda L / 2}{r_A} \frac{\sin \left(1-r_A\right) \lambda L / 2}{\left(1-r_A\right)} \sin \frac{\lambda L}{2}\right] \\
& -8 s_{13} s_{23}^2\left(\left|\varepsilon_{e \mu}\right| s_{23} c_X+\left|\varepsilon_{e\tau}\right| c_{23} c_{\omega}\right) r_A\left[\frac{\sin ^2\left(1-r_A\right) \lambda L / 2}{\left(1-r_A\right)^2}\right] \text ,
\end{aligned}
\end{equation}

\noindent where $r_A = A(x)/\Delta m^{2}_{31} $, $\lambda = \Delta m^{2}_{31}/2E$, $X = \delta_{e\mu}+\delta_{CP}$, $\omega = \delta_{e\tau}+\delta_{CP}$ being $\delta_{\alpha \beta}$ the complex phase of the NSI couplings, and $s_{ij}, \  c_{ij}, \ s_{2\times ij}, \  c_{2\times ij}  = \sin\theta_{ij}, \ \cos \theta_{ij}, \ \sin 2\theta_{ij}, \ \cos 2\theta_{ij}$. The coupling $\varepsilon_{ee}-\varepsilon_{\mu\mu}$, which is neglected in the analysis, does not appear at leading order in the two oscillation channels to which the experiment is most sensitive.

\subsection{NSI expectation in the $\nu_\mu - \nu_\tau$ sector}
\label{sec:NSImutau}
%% This sentence needs reformulation.
% Event samples in the signal region of ORCA6 are mostly dominated by $\nu_\mu / \bar{\nu}_{\mu}-$~CC interactions, for which reason the discussion on the NSI effects will mainly focus on the muon neutrino survival probabilities shown in Figs. \ref{fig:NuMuProbs} and \ref{fig:NuMuOsc}. 
Event samples rich in $\nu_\mu / \bar{\nu}_{\mu}$ CC interactions above 10 GeV are well described within the two-flavour oscillation picture, for which reason the discussion on the NSI effects in the $\nu_\mu - \nu_\tau$ sector will mainly focus on the muon neutrino survival probabilities shown in Figs.~\ref{fig:NuMuProbs} and \ref{fig:NuMuOsc}.

The NC flavour-violating $\varepsilon_{\mu\tau}$ and non-universality $\varepsilon_{\tau\tau} - \varepsilon_{\mu\mu}$ dominate this channel over the remaining NSI couplings to leading order in $\sqrt{\Delta m^{2}_{21}/\Delta m^{2}_{31}} \sim \sqrt{|\varepsilon_{\alpha \beta}|} \sim \sin \theta_{13}$ (see eq.~\eqref{eq:NSImumu}) \cite{Wang:2018dwk, Chatterjee:2014gxa}, providing the highest sensitivity to any of the parameters in neutrino telescopes.   

The top-left panel in Fig.~\ref{fig:NuMuProbs} shows the effect of real-valued $\varepsilon_{\mu\tau} = \pm 7.5 \times 10^{-3}$ at oscillation probability level. Consistent with the derivations in \cite{Esmaili:2013fva}, the mentioned coupling shifts the position of the oscillation minimum present above 20 GeV by modifying the effective value of $\Delta m^{2}_{31}$ in matter. This effect can, however, be washed out by an exact $\mu-\tau$ maximal mixing, namely $\theta_{23}=45^\circ$ \cite{Esmaili:2013fva}. Additionally, above $\sim$ 60 GeV both signs of $\varepsilon_{\mu\tau}$ exhibit an oscillation enhancement compared to standard matter effects, which is sustained across many decades of energy as shown by the blue areas in the top-left oscillogram of Fig.~\ref{fig:NuMuOsc}. This signature would lead to a net $\nu_{\mu}+\bar{\nu}_\mu$ deficit expected in the high-energy range after convolution with the atmospheric flux and detector response. Neutrinos in this regime tend to be reconstructed with lower energies due to finite detector resolution, entering the relevant range for the present study and therefore enhancing the sentivity to $\varepsilon_{\mu\tau}$. 

The modification of the oscillation pattern due to a non-zero value of $\varepsilon_{\tau\tau} - \varepsilon_{\mu\mu}$ is shown in the top-right panel of Fig.~\ref{fig:NuMuProbs}. Although appearing at the same order as $\varepsilon_{\mu\tau}$, the oscillation terms driven by $\varepsilon_{\tau\tau} - \varepsilon_{\mu\mu}$ are suppressed in the vicinity of maximal mixing by a factor $\cos 2\theta_{23}$ \cite{Chatterjee:2014gxa, Kikuchi:2008vq}, as seen in the second line of eq.~\eqref{eq:NSImumu}, consequently relaxing the sensitivity by at least an order of magnitude with respect to $\varepsilon_{\mu\tau}$. Above 10 GeV and for the most up-going trajectories, the effect arises as a damping of the oscillation probabilities around the oscillation dip, being highly degenerate with respect to the sign of $\varepsilon_{\tau\tau} - \varepsilon_{\mu\mu}$. The degeneracy can become almost exact if $\theta_{23}=45^\circ$.

\subsection{NSI expectation in the $\nu_e$ sector}
\label{sec:NSIelec}

The NSI couplings involving the electron flavour $\varepsilon_{e\tau}$ and $\varepsilon_{e\mu}$ appear at subleading order in the $\nu_\mu / \bar{\nu}_\mu$ survival probabilities from eq.~\eqref{eq:NSImumu} \cite{Kikuchi:2008vq}, rendering neutrino telescopes far less sensitive to them compared to the NSI in the $\mu - \tau$ sector. Their effects can still be visualised for large enough couplings in the bottom panels of Fig.~\ref{fig:NuMuProbs}. The modifications due to $\varepsilon_{e\tau}$ approximately resemble that of $\varepsilon_{\tau\tau} - \varepsilon_{\mu\mu}$ at the oscillation valley, but the similarity breaks down below 10 GeV. As opposed to $\varepsilon_{e\tau}$, subtle high-energy effects are expected in the presence of non-zero $\varepsilon_{e\mu}$ independently of its sign, as shown in the bottom-right panel of Fig.~\ref{fig:NuMuProbs}.

The parameters in the $e-\tau$ and $e-\mu$ sectors lead the NSI modifications at first order in the $\nu_\mu \rightarrow \nu_e$ transitions seen in eq.~\eqref{eq:NSImue}, whereas the couplings $\varepsilon_{\mu\tau}$ and  $\varepsilon_{\tau\tau} - \varepsilon_{\mu\mu}$ appear at second order in the expansion, thus to a good approximation they decouple from the former \cite{Wang:2018dwk, Kikuchi:2008vq}. As a result, the electron appearance probability is only subtly modified in the top panels of Fig.~\ref{fig:NuEProbs}, for the same choice of parameters as in Fig.~\ref{fig:NuMuProbs}, while the effects expected from $\varepsilon_{e\tau}$ and $\varepsilon_{e\mu}$ are most prominent in this channel. For this reason, atmospheric neutrino samples rich in $\nu_e/\bar{\nu}_e$ CC interactions enhance the sensitivity of the detector to these couplings by enabling a precise measurement of electron neutrino and antineutrino appearance. In Fig.~\ref{fig:NuMuOsc}, the NSI signatures in the $e-\mu$ and $e-\tau$ sectors still show significant modifications of oscillation probabilities for baselines with $\cos\theta_z > -0.8$, in contrast to the mostly up-going NSI in the $\mu-\tau$ sector, due to the periodic functional dependence on the matter potential entering through $r_A$ inside the brackets of eq.~\eqref{eq:NSImue}.  

%%%%%%%%%%%%%%%%%%%%%%%%%%%%%%%%%%%%%%%%%%%%%%%%%%%%%%%%%%%%%%%%%%%%%%%%%%%%%%%%%%%%%%%%%%%%%%%%%%%%%%%%%%%%%%%%%%%%%%%%%%%%%%%
%\newpage
%\input{detector.tex}
\section{The KM3NeT/ORCA detector}
\label{sec:detector}

The KM3NeT Collaboration  is building two water Cherenkov neutrino detectors pursuing different physics goals  but sharing a common technology \cite{KM3Net:2016zxf}. The infrastructure is distributed between two sites in the Mediterranean Sea, one located 100 km offshore Portopalo di Capo Passero (Sicily, Italy) with KM3NeT/ARCA  (Astroparticle Research with Cosmics in the Abyss) at a depth of 3500 m, and the other one hosting the KM3NeT/ORCA detector (Oscillation Research with Cosmics in the Abyss), 40 km offshore Toulon (France) at 2450 m below sea level.  While ARCA's sparse layout is optimised for TeV$-$PeV astrophysical neutrino detection, ORCA is more densely instrumented in order to study the oscillations of $\mathcal{O}$~(1$-$100~GeV) atmospheric neutrinos, produced in air showers induced by cosmic ray interactions in the atmosphere. %The ultimate goal of ORCA is the precision measurement of atmospheric neutrino mixing parameters and the determination of the neutrino mass ordering.  
% we said this laready before I think. SHould be repeated?

Neutrino interactions near or inside the instrumented volume produce secondary charged particles, whose Cherenkov light yield is detected by a three-dimensional grid of Digital Optical Modules (DOMs) \cite{KM3NeT:2022pnv} arranged on vertical detections units \cite{adrianmartinez:in2p3-01220200} in order to reconstruct the parent neutrino energy and direction. The DUs are mechanical structures anchored to the seabed, held vertically by a submerged buoy at the top  and the buoyancy of the 18 DOMs, each housing 31 photomultiplier tubes (PMTs).  Additionally, the DOMs contain the readout electronics and all the sensors necessary for the positioning, orientation and time calibration of the detector \cite{adrianmartinez:in2p3-01220200, Real:2019rcv}. ORCA and ARCA are currently undergoing a modular construction, following a horizontal layout with  20 m average horizontal spacing between neighbouring DUs in the former, and 90 m in the latter. The DOMs are positioned every 9 m vertically along the DUs in ORCA and 36 m in ARCA. The final configuration foreseen for ORCA will host 115 DUs while twice as many will constitute ARCA. 

%%%%%%%%%%%%%%%%%%%%%%%%%%%%%%%%%%%%%%%%%%%%%%%%%%%%%%%%%%%%%%%%%%%%%%%%%%%%%%%%%%%%%%%%%%%%%%%%%%%%%%%%%%%%%%%%%%%%%%%%%%%%%%%%%%%%%%%
%\input{sample_selection.tex}
\section{Event sample and selection}
\label{sec:dselection}

Starting in January 2020, the configuration of ORCA with the first six DUs, ORCA6, uninterruptedly took data until November 2021, when the detector was expanded with more DUs in successive deployment campaigns. The dataset used in this work covers 433 kton-years of exposure of ORCA6, which were selected according to strict quality criteria on the environmental conditions and stability of the data taking. 

The ground-level event filtering is designed to cope with the background dominated by optical noise. Situated in the depths of the Mediterranean Sea, the detector is sensitive to the constant baseline of optical photons coming from the decay chain of $^{40}$K, as well as the seasonally-dependent bioluminescence from microorganisms. The multi-PMT configuration is exploited by online triggers in order to retain the causally correlated optical-photon arrivals named \textit{hits}, but a non-negligible burden of random coincidences still survive at this level. Cuts are set based on quality outputs from event reconstruction algorithms and the number of triggered hits per event, in order to reduce the optical noise contamination of the sample below the percent level.

Atmospheric muons are produced together with neutrinos in air showers, and they outnumber the latter by over seven orders of magnitude at the present selection stage. An up-going reconstructed direction is required for events to be selected, as muons are not able to cross the Earth. Down-going muons with faint Cherenkov light yield can be reconstructed as up-going, hence a dedicated Boosted Decision Tree (BDT) \cite{KM3NeT:2024ecf} is employed to discriminate between the former and the signal of neutrino interactions. The BDT is trained on features engineered from the hit distribution and outputs from dedicated reconstruction algorithms. A cut on the BDT score reduces the muon contamination down to 2\% while keeping 60\% of the estimated neutrino signal at the previous selection stage.

The last selection level is aimed at distinguishing the two main neutrino interaction signatures. The event topology denoted as track-like originates mainly from the $\nu_\mu / \bar{\nu}_{\mu}$ CC interactions near or inside the instrumented volume, whose outgoing muon is able to travel from tens to hundreds of metres in seawater leaving a characteristic yield of Cherenkov photons. In contrast, the shower-like topology corresponds to a more isotropic photon distribution in the detector, associated with neutrino interaction channels where only hadronic and electromagnetic cascades are produced, with a limited spatial extension compared to the track topology. This includes CC interactions of $\nu_e / \bar{\nu}_{e}$ and $\nu_\tau/ \bar{\nu}_{\tau}$ with non-muonic decays, as well as NC interactions of all neutrino flavours.

% owing to the limited spatial extension of the hadronic interaction vertex and the electromagnetic showers induced by outgoing electrons in the $\nu_e / \bar{\nu}_{e}$ CC interactions. 

\begin{table}[t!]
\centering
\begin{tabular}{c|ccc|c}
\hline
          \textbf{Channel}   & \textbf{HP Tracks} & \textbf{LP Tracks} & \textbf{Showers} & \textbf{Total} \\ \hline \hline
$\nu_\mu$ CC                   & 1166.2               & 1187.1              & 670.2              & 3023.5           \\
$\bar{\nu}_\mu$ CC             & 612.4                & 600.8               & 236.0              & 1449.2           \\
$\nu_\mu+\bar{\nu}_\mu$ CC     & 1778.6               & 1787.9               & 906.2              & 4472.7           \\ \hline
$\nu_e$ CC                     & 36.9                 & 62.1                & 434.5              & 533.5            \\
$\bar{\nu}_e$ CC               & 14.0                 & 22.9                 & 172.5              & 209.4            \\
$\nu_e + \bar{\nu}_e$ CC       & 50.9                 & 85.0                 & 607.0              & 742.9            \\ \hline
$\nu_\tau$ CC                  & 14.0                 & 13.0                 & 95.3               & 122.3            \\
$\bar{\nu}_\tau$ CC            & 6.4                  & 5.7                  & 37.3               & 49.4             \\
$\nu_\tau + \bar{\nu}_\tau$ CC & 20.4                 & 18.7                 & 132.6              & 171.7            \\ \hline
$\nu$ NC                       & 9.6                 & 16.9                 & 224.3              & 250.8           \\
$\bar{\nu}$ NC                 & 3.0                  & 5.1                  & 66.7               & 74.8             \\
$\nu+\bar{\nu}$ NC             & 12.6                 & 22.0                 & 291.0              & 325.6            \\ \hline
Atm. Muons                     & 7.1                  & 87.5                 & 22.2               & 116.8            \\ \hline
\textbf{Total MC}              & 1869.6               & 2001.1               & 1959.0             & 5829.7           \\ \hline \hline
\textbf{Total Data}            & 1868               & 2002               & 1958             & 5828           \\ \hline
\end{tabular}
\caption{Expected number of events at the best fit of oscillation and nuisance parameters under the null hypothesis (standard interactions), broken down by neutrino flavour and interaction type, as well as the remaining atmospheric muon background. The contributions are further divided into the three classification samples used in the analysis: high-purity tracks (HP), low-purity tracks (LP) and showers. Table taken from \cite{KM3NeT:2024ecf}.}
\label{tab:samplecomp}
\end{table}

A second BDT is trained to discriminate between the two types of signatures. Combining the two BDT outputs by employing non-overlapping BDT score ranges, the dataset comprising 5828 observed events is split into three classes: high-purity tracks (HP) with 0.3\% atmospheric muon contamination and estimated 95\% $\nu_{\mu}$ CC purity, low-purity (LP) tracks with 4\% muon contamination and 90\% $\nu_{\mu}$ CC purity, and a shower class with 46\% of muon neutrinos mixed with the remaining neutrino flavours and NC interactions. The specific BDT cuts applied for this event classification into the sets were optimised to provide the highest sensitivity to neutrino oscillation parameters, and were identically chosen for all ORCA analyses performed on the 433 kton-years dataset \cite{KM3NeT:2024ecf}. The detailed composition of the analysis classes is shown in Table \ref{tab:samplecomp}. 

The high- and low-purity distinction for tracks stems from the benefit of isolating events with better angular resolution from less accurately reconstructed ones, since the BDT scores serve as a quality parameter. For the shower channel such separation is not feasible, given the reduced statistics in that class compared to the two track channels combined.

While the three classes are approximately equally populated, the high-purity track class enhances the $\nu_{\mu}$-disappearance signal rendering it the most sensitive to $\varepsilon_{\mu\tau}$ and  $\varepsilon_{\tau\tau} - \varepsilon_{\mu\mu}$, whereas the shower class opens the possibility to effectively constrain the electron NSI sector, namely $\varepsilon_{e\tau}$ and $\varepsilon_{e \mu}$.

%%%%%%%%%%%%%%%%%%%%%%%%%%%%%%%%%%%%%%%%%%%%%%%%%%%%%%%%%%%%%%%%%%%%%%%%%%%%%%%%%%%%%

\newpage

\section{Analysis method}
\label{sec:analysis}

The analysis proceeds by comparing the observation with Monte Carlo (MC) simulated templates weighted according to the hypothesis being tested. The generated events are mapped from the simulation phase space to the templates, binned in reconstructed $\log(E/\mathrm{GeV})$ and $\cos \theta_z$, by means of a detector response modelled from the MC simulation itself, which convolutes the total detection, reconstruction and classification efficiency within a particular ($\log(E/\mathrm{GeV}), \cos \theta_z$) bin in the reconstruction space. An overview of the ORCA MC simulation chain and response matrix is presented in \cite{KM3NeT:2021ozk, KM3NeT:2021uez, KM3NeT:2023ncz, KM3NeT:2024ecf}.

A Maximum Likelihood Estimation (MLE) is performed in order to extract the NSI and nuisance parameters which best fit the observed distributions, under the hypothesis of choice. A binned negative Poisson log-likelihood is used as 

\begin{multline}\label{eq:likelihood}
    -2\log \mathcal{L} = \sum^{n_{\mathrm{rec}}}_{i, j}2\left[\beta_{ij} N_{i j}^{\mathrm{M}}(\vec{\omega}, \vec{\eta})-N_{i j}^{\mathrm{D}}+N_{i j}^{\mathrm{D}} \log \left(\frac{N_{i j}^{\mathrm{D}}}{\beta_{ij}N_{i j}^{\mathrm{M}}\left(\vec{\omega}, \vec{\eta} \right)}\right)\right] + \frac{(\beta_{ij}-1)^2}{\sigma^{2}_{\beta,  i j}} \\
 +\sum_{k}\left(\frac{\eta_k-\left\langle\eta_k\right\rangle}{\sigma_k}\right)^2 \text ,
\end{multline} 

\noindent which is expected to asymptotically follow a $\chi^2$-distribution \cite{BAKER1984437}. The sum runs over the $n_{\mathrm{rec}}$ bins in the reconstructed energy and cosine of the zenith angle, $N^{\mathrm{M}}_{ij}$ and $N^{\mathrm{D}}_{ij}$ denote the model prediction and data recorded for the energy bin $i$ and cosine of the zenith bin $j$, $\vec{\omega}$  is the set of parameters of interest (PoI) and $\vec{\eta} $ are the nuisance parameters. The likelihood from eq.~\eqref{eq:likelihood} is further corrected by the second term due to the finiteness of the available MC statistics, following Conway's approximation \cite{Conway:2011in} of the original Beeston-Barlow method \cite{Barlow:1993dm}. The parameters $\beta_{ij}$ model the bin-by-bin uncorrelated  MC uncertainty and act as nuisance parameters, assumed to be normally distributed $\beta_{ij}\sim \mathcal{N}(1,\sigma^{2}_{\beta, i j})$ in each bin, where $\sigma_{\beta, i j}$ is the uncertainty on the bin content obtained as the error of the sum of event weights. The $\beta_{ij}$ parameters are independent of the PoIs and other nuisance parameters and are analytically computed as explained in \cite{KM3NeT:2024ecf}.  Finally, the third term of the likelihood accounts for external constraints on nuisance parameters which are assumed to be Gaussian.
%and is evaluated as the error on the sum of weights of events reconstructed in that bin

Fourteen bins in reconstructed $\log(E/\mathrm{GeV})$ were set from 2 GeV up to 100 GeV for high- and low-purity tracks, while an extra bin up to 1 TeV was added for the shower class. Ten bins are linearly distributed in $\cos \theta_z$ between $[-1,0]$. The binning was optimised to provide enough statistics in otherwise sparse corners of the phase space, ensuring sufficient bin-by-bin populations for the correct behaviour of finite-MC corrections. 

In order to construct confidence level intervals for the NSI coupling strengths, the log-likelihood ratio is chosen as test statistic (TS):

\begin{equation}\label{eq:LLR}
    -2\Delta \log \mathcal{L}(\vec{\omega}_0) = -2\log \left(\frac{\mathcal{L}(\vec{\omega}_0, \hat{\hat{\vec{\eta}}})}{\mathcal{L}(\hat{\vec{\omega}},\hat{\vec{\eta}})}\right) \text ,
\end{equation}

\noindent where $\vec{\omega}_0$ is a specific point within the specified grid of NSI parameters, $\hat{\hat{\vec{\eta}}}$ is the MLE of the nuisance parameters given the fixed $\vec{\omega}_0$, $\hat{\vec{\omega}}$ is the global MLE of the NSI coupling and $\hat{\vec{\eta}}$ is that of the nuisance parameters. Grids assumed for different NSI hypotheses would then be scanned successively at different values of $\vec{\omega}_0$, constructing the $-2\Delta \log \mathcal{L}$ profiles and contours as specified by eqs.~\eqref{eq:LLR} and \eqref{eq:likelihood}. The allowed region of each NSI hypothesis at a given confidence level is obtained by comparing the values of eq.~\eqref{eq:LLR} with the corresponding quantiles of a $\chi^2$-distribution with one or two degrees of freedom assuming Wilks' theorem~\cite{Wilks:1938dza}. % and in the case of one-dimensional likelihood scans, a second complementary confidence region is obtained by directly sampling the underlying test-statistic distribution, following the Feldman-Cousins (FC) prescription \cite{PhysRevD.57.3873}.  

\begin{table}[t!]
\centering
\renewcommand{\arraystretch}{1.1} % Adjust the row spacing (1.5 times the default)
\begin{tabular}{lcc}
\hline
\multicolumn{1}{c}{\textbf{Parameter}}        & \multicolumn{1}{l}{}   & \textbf{Uncertainty} \\ \hline \hline
\multicolumn{2}{c}{Oscillations:}                                      & \multicolumn{1}{l}{} \\
\multicolumn{2}{c}{$\Delta m^{2}_{31}$} & ---                  \\
\multicolumn{2}{c}{$\theta_{23}$}                           & ---                  \\ \hline
\multicolumn{2}{c}{Normalisations \& cross sections:}                  & \multicolumn{1}{l}{} \\
High-purity tracks                            & $f_{\mathrm{HPT}}$     & ---                  \\
Showers                                       & $f_{\mathrm{S}}$       & ---                  \\
Overall                                       & $f_{\mathrm{all}}$     & ---                  \\
Atm. muon background                          & $f_{\mathrm{\mu}}$     & ---                  \\
NC normalisation                              & $f_{\mathrm{NC}}$      & $\pm 20$\%           \\
$\nu_\tau / \bar{\nu}_\tau$ CC normalisation  & $f_{\mathrm{\tau CC}}$ & $\pm 20$\%           \\ \hline
\multicolumn{2}{c}{Detector:}                                          & \multicolumn{1}{l}{} \\
High-energy light yield                       & $f_{\mathrm{HE}}$      & $\pm 50$\%           \\
Energy scale                                  & $E_s$                  & $\pm 9$\%            \\ \hline
\multicolumn{2}{c}{Atmospheric neutrino flux:}                         & \multicolumn{1}{l}{} \\
Spectral index                                & $\delta_\gamma$        & $\pm 0.3$            \\
Ratio up-going to horizontal $\nu$            & $\delta_\theta$        & $\pm 2$\%            \\
$\nu_{\mu}/\bar{\nu}_{\mu}$ ratio             & $s_{\mu\bar{\mu}}$     & $\pm 5$\%            \\
$\nu_{e}/\bar{\nu}_{e}$ ratio                 & $s_{e\bar{e}}$         & $\pm 7$\%            \\
$\nu_{\mu}/\nu_{e}$ ratio                     & $s_{e\mu}$             & $\pm 2$\%            \\ \hline
\end{tabular}
\caption{Nuisance parameters and their corresponding uncertainties used as prior widths in the analysis, taken from \cite{KM3NeT:2024ecf}. }
\label{tab:nuisance}
\end{table}

\subsection{Nuisance parameters}
\label{sec:systematics}

Fifteen nuisance parameters are considered in this study modelling a variety of systematic uncertainties (see Table \ref{tab:nuisance}), as was done in \cite{KM3NeT:2024ecf}. The standard oscillation hypothesis being nested within the NSI scenario requires that $\Delta m^{2}_{31}$ and $\theta_{23}$ are profiled over with no assumed prior, and the central values taken from NuFit 5.0 \cite{Esteban:2020cvm}. The remaining oscillation parameters were kept fixed as their impact was found to be negligible in this analysis. The MLE estimators are computed by performing fits with eight different combinations of starting points: normal and inverted mass ordering $\Delta m^{2}_{31}=\{ 2.517, -2.428\}\times 10^{-3} \ \mathrm{eV^2}$, $\theta_{23}$-value within the upper and lower octant $\theta_{23} = \{ 40^\circ, 50^\circ \}$, and finally a starting point of the energy scale 5\%~larger and smaller than nominal, $E_s = \{ 0.95, 1.05\}$, following the observation of degenerate minima of this systematic. In order to avoid falling into a local minimum of the negative log-likelihood function, only the global minimum of eq.~\eqref{eq:likelihood} is extracted out of the 8 settings.

Three normalisation factors account for the uncertainty inherent to the event classification into three sets, whereas an additional systematic models the total yield of the sparse atmospheric muon background which enters the event selection. Total inclusive yield uncertainties are considered in the all-flavour NC and $\nu_\tau$ CC interactions. All normalisation, cross section and background systematics listed in Table 2 act as scaling factors accounting for selection efficiency,
mismodelling and cross-section uncertainties that modify the total number of events expected in the corresponding event class or type. The prior widths assumed for systematics modelling the detector response are based on dedicated MC studies. Five nuisance parameters take into account the uncertainties on the atmospheric neutrino flux, two of them representing its spectral index and zenith-angle dependence \cite{Honda:2006qj}, and the remaining three implemented as flavour ratio uncertainties as suggested in \cite{Barr:2006it}. The baseline neutrino flux used in this analysis is the azimuth-averaged HKKMS 2015 model \cite{PhysRevD.92.023004} computed at the Fréjus site without mountain over the detector. A detailed discussion of the implementation of the uncertainties is covered in \cite{KM3NeT:2024ecf}.

% Additionally, a normalization systematic acting on neutrinos above 500 GeV in true energy is implemented, in order to account for the uncertainty in the high energy light simulation by JSirene\footnote{It is believed that the very high energy neutrinos ($E_{true}\geq 500$ GeV), simulated by JSirene in our MC, are the main source driving a characteristic $\sim 5\%$ event deficit of the MC respect to data. Such normalization then becomes meaningful. For implementation details, see \href{https://indico.cern.ch/event/1282683/contributions/5388908/attachments/2643799/4575810/OscWG_Intro_20230509.pdf}{Joao Coelho's presentation}. }. On top of that,

%%%%%%%%%%%%%%%%%%%%%%%%%%%%%%%%%%%%%%%%%%%%%%%%%%%%%%%%%%%%%%%%%%%%%%%%%%%%%%%%%%%%%%%%%%%%%%%%%%%%%%%%%%%%%%%%%%%%%%%%%%%
%\input{results.tex}
\section{Results}
\label{sec:results}

The different NSI scenarios are considered by assuming only one coupling strength to be non-zero at a time, in order to keep the analysis feasible within the available computational resources. This procedure naturally neglects any correlation between couplings which can lead to partial cancellation of effects, and ultimately to relaxed bounds on the parameters compared to the one-by-one case. 

% The most general approach should include all possible correlations \cite{Coloma:2023ixt} and will be followed in future analyses. 

Section \ref{sec:bestfit} presents the values of the NSI couplings which best fit the observed data. Sections \ref{sec:mutausector} and \ref{sec:elecsector} present the allowed regions for the flavour-violating and non-universality NSI couplings obtained from the data. In the former case, the complex-valued nature of off-diagonal parameters requires log-likelihood ratio scans as functions of the correlated complex phase and modulus. From these, the allowed regions in $(|\varepsilon_{\alpha\beta}|,\delta_{\alpha\beta})$ are extracted assuming a $\chi^2$-behaviour of the TS following Wilks' theorem. One-dimensional scans of the TS are provided for the diagonal coupling $\varepsilon_{\tau\tau}-\varepsilon_{\mu\mu}$, and additionally for the off-diagonal couplings by fixing the complex phase to $\delta_{\alpha\beta}=0$ or $ \pi$, as was customarily done by other experiments. 

% 1D scans are provided together with the FC corrections to the likelihood ratio, extracted as the 68\% and 90\% quantiles of the TS distributions sampled from pseudo-experiments generated at the best fit of each NSI grid point, following the "Profiled Feldman-Cousins" approach proposed in ref. \cite{NOvA:2022wnj}.

% \begin{table}[t!]
% \centering
% \begin{tabular}{ccc}
% \hline
% \textbf{Hypothesis}                           & \textbf{\begin{tabular}[c]{@{}c@{}}Best fit\\ $|\varepsilon_{ij}|, \ \delta_{ij}$\end{tabular}}       & \textbf{$-2\log(\mathcal{L}_{\mathrm{SI}}/\mathcal{L}_{\mathrm{NSI}}) $}  \\ \hline \hline
% $\varepsilon_{\mu\tau}$                       & $0.01^{+0.03}_{-0.01}\times 10^{-1}, \ {{0 }^{+360^\circ}_{-0^\circ}}$       & 0.15           \\ %0.92
% $\varepsilon_{\tau\tau}-\varepsilon_{\mu\mu}$ & $0.00 \pm 0.01$, ---         & 0.01       \\ %0.92
% $\varepsilon_{e\tau}$                         & $0.04^{+0.02}_{-0.04}, \ {200}^{+80^\circ}_{-100^\circ}$ & 1.02        \\%0.47
% $\varepsilon_{e\mu}$                          & $0.03^{+0.02}_{-0.03}, \ {130}^{+90^\circ}_{-70^\circ}$ & 1.24           \\ \hline \hline %0.42
% \end{tabular}
% \caption{Results for the NSI couplings from the best fits to the observed data. }
% \label{tab:bestfits}
% \end{table}

\subsection{Best fit}
\label{sec:bestfit}

Table \ref{tab:bestfits} presents the best fit obtained under each NSI hypothesis alone.  No significant deviation from the standard interaction case was found in any of the fits to the data, as evidenced by the low log-likelihood ratios with respect to the SI hypothesis, where the NSI fit had 2 (1) fewer degrees of freedom for complex (real) couplings. The goodness-of-fit p-value resulted in $\sim$1.2~\% for the four scenarios.
\\

\begin{table}[h!]
\centering
\begin{tabular}{cccc}
\hline
\multirow{2}{*}{\textbf{NSI coupling}}          & \multicolumn{2}{c}{\textbf{Best fit}}                                     & \multirow{2}{*}{\textbf{$-2\log(\mathcal{L}_{\mathrm{SI}}/\mathcal{L}_{\mathrm{NSI}}) $}} \\
                                              & \textbf{$|\varepsilon_{\alpha\beta}|$} & \textbf{$\delta_{\alpha\beta}$}  &                                                                                           \\[2.5pt] \hline 
                                              \\[-5pt]
                                             
$\varepsilon_{\mu\tau}$                       & $0.01^{+0.03}_{-0.01}\times 10^{-1}$    & ${{0 }^{+360^\circ}_{-0^\circ}}$ & 0.15                                                                                      \\[1.5pt]
$\varepsilon_{\tau\tau}-\varepsilon_{\mu\mu}$ & $0.00 \pm 0.01$                        & ---                              & 0.01                                                                                      \\[2.5pt]
$\varepsilon_{e\tau}$                         & $0.04^{+0.02}_{-0.04}$                 & ${200}^{+80^\circ}_{-100^\circ}$ & 1.02                                                                                      \\[2.5pt]
$\varepsilon_{e\mu}$                          & $0.03^{+0.02}_{-0.03}$                 & ${130}^{+90^\circ}_{-70^\circ}$  & 1.24
                                                \\[-5pt]
                                                \\ \hline
\end{tabular}
\caption{Results for the NSI couplings from the best fits to the observed data.}
\label{tab:bestfits}
\end{table}

\begin{figure}[t!]
\centering
\includegraphics[width=0.75\textwidth]{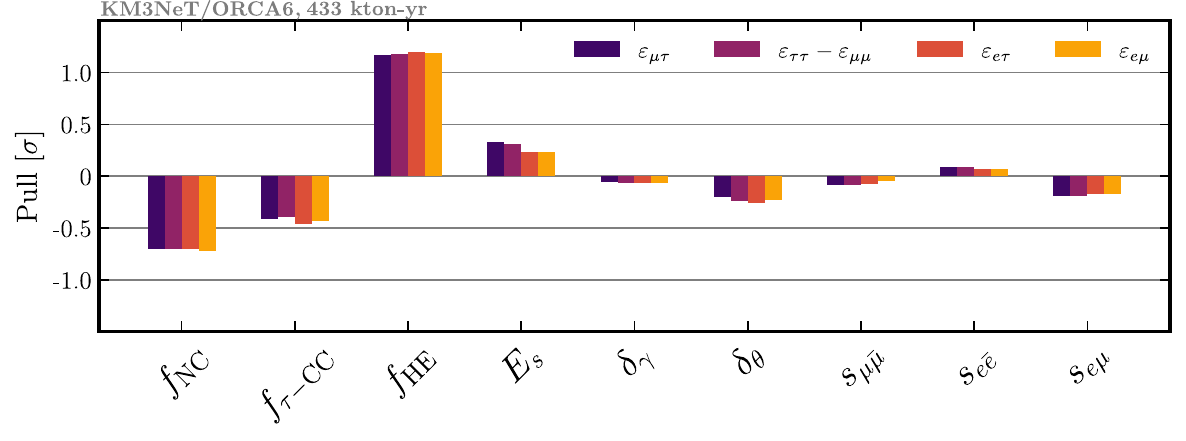}
\qquad
\caption{Statistical pull on the nuisance parameters with prior constraints obtained in the four fits to the data. The explanation of the parameters is covered in Table \ref{tab:nuisance}. }
\label{fig:pulls}
\end{figure}

The statistical pull on the nuisance parameters with external priors are shown in Fig.~\ref{fig:pulls} for the four NSI hypotheses. None of the constrained systematics exhibits a statistically relevant pull. The oscillation parameters $\Delta m^{2}_{31}$ and $\theta_{23}$ were fitted within 4\% of the standard interaction best fit from \cite{KM3NeT:2024ecf}, with the only exception of $\Delta m^{2}_{31}$ being 10\% away under the $\varepsilon_{e\tau}$-only hypothesis. The detailed impact of the nuisance parameters on each NSI coupling is presented in appendix \ref{app:nuisance}, Figs.~\ref{fig:shifts1} and \ref{fig:shifts2}. The oscillation parameters $\theta_{23}$ and $\Delta m^{2}_{31}$ exhibit the highest influence due to the degeneracy produced on the oscillation minimum by varying them and the NSI couplings.

%% 
%% 
%\newpage

\begin{figure}[h!]
\centering
\includegraphics[width=0.85\textwidth]{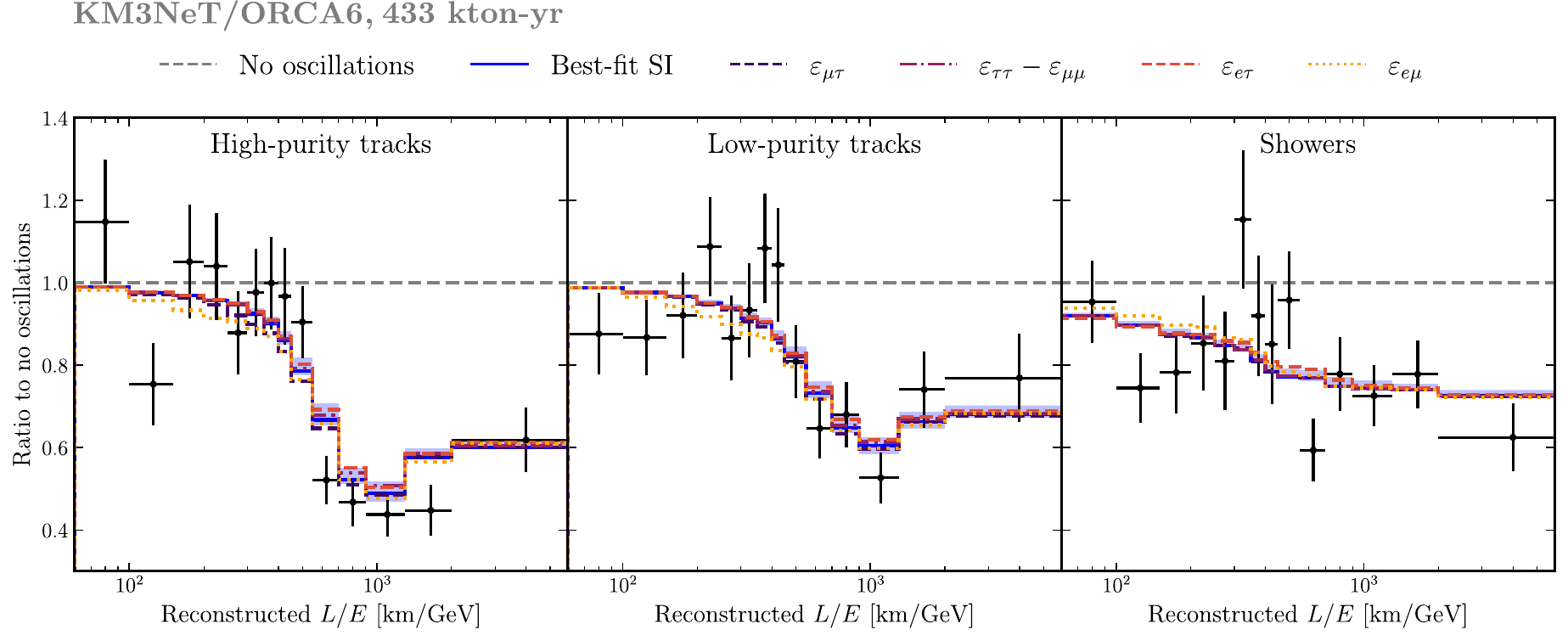}
%\qquad
\caption{Ratio of events with respect to the non-oscillation hypothesis as a function of the reconstructed baseline over energy, $L/E$. The data points are shown together with the prediction from the standard interaction (SI) best fit and four different NSI scenarios, where the corresponding coupling strength is fixed to a value rejected at 90\% CL while keeping the nuisance parameters at the best fit. The shaded bands cover 68\% of the trials drawn by fitting the SI hypothesis to 1000 pseudo-experiments generated from Poisson fluctuations of the data best-fit MC template. Observation and expectations are separated into the three event sets: high-purity tracks (left), low-purity tracks (centre) and showers (right). }
\label{fig:LoE}
\end{figure}

Fig.~\ref{fig:LoE} shows the ratios of events with respect to the non-oscillation hypothesis for the observed data and four NSI scenarios. The choice to display NSI coupling values rejected at 90\% CL allows for highlighting the NSI phenomenology, which would otherwise be washed out at their best-fit values since no statistically significant pull was observed on any NSI parameter.

%\newpage

% \begin{figure}[h!]

% \centering

% \subfloat[]{
% \centering
% \hspace{-1.7cm}

% \label{fig:LoEtracks}
% \includegraphics[height=6.5cm]{imgs/results/LoE_tracks_cut_-0.000000.pdf}
% }
% \subfloat[]{
% \centering
% \label{fig:LoEtracks_c}
% \includegraphics[height=6.5cm]{imgs/results/LoE_tracks_c_cut_-0.000000.pdf}} \
% \subfloat[]{
% \centering
% \label{fig:LoEshowers}
% \includegraphics[height=6.5cm]{imgs/results/LoE_showers_cut_-0.000000.pdf}}

% \caption{Ratio of events with respect to the no-oscillation hypothesis as a function of the baseline over reconstructed energy bin ($L/E$). The data points are shown together with the prediction from the standard interaction best fit and four different NSI scenarios, where the corresponding coupling strength is fixed to a value rejected at 90\% CL while keeping the nuisance parameters at the best fit. The shaded bands cover 68\% of the trials drawn by fitting the standard interaction hypothesis to 1000 pseudo-experiments, generated from Poisson fluctuations of the data best-fit MC template. Observation and expectations are separated into the three event sets: high-purity tracks (\ref{fig:LoEtracks}), low-purity tracks( \ref{fig:LoEtracks_c}) and showers (\ref{fig:LoEshowers}). }
% \label{fig:LoE}
% \end{figure}

\newpage
\subsection{NSI in the $\nu_{\mu}-\nu_{\tau}$ sector}
\label{sec:mutausector}

\begin{figure}[h!]
\centering
%\subfloat[$\varepsilon_{\mu\tau}$]{
\centering

\includegraphics[height=7.4cm]{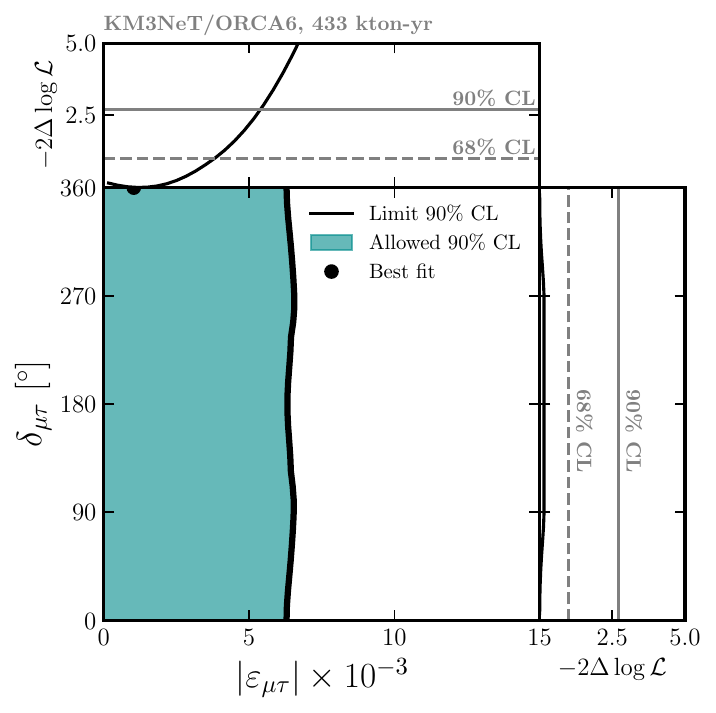}
%}
% \subfloat[$\varepsilon_{e\tau}$]{
% \centering
% \label{fig:complexnsib}
% \includegraphics[height=6.3cm]{imgs/results/contour_eps_et_delta_et_paper_color.pdf}} \
% \subfloat[$\varepsilon_{e\mu}$]{
% \centering
% \label{fig:complexnsic}
% \includegraphics[height=6.3cm]{imgs/results/contour_epm_em_delta_em_paper_color.pdf}}

\caption{Observed log-likelihood ratio contour at 90\% CL of the complex, flavour-violating NSI coupling $\varepsilon_{\mu\tau}$. The shaded area indicates the allowed region of the parameter space $\{|\varepsilon_{\mu\tau}|,\delta_{\mu\tau}\}$ at the reported CL, containing the best-fit point (dot). The top and side panels show the projections of $-2\Delta \log \mathcal{L}$ when the other variable is profiled over. The dashed and solid grey lines of the adjoint projections are the 68\% and 90\% CL of the $\chi^2$ distribution with one degree of freedom.}
%\label{fig:complexnsia}
\label{fig:complexnsia}
\end{figure}

Fig.~\ref{fig:complexnsia} presents the allowed region observed for the correlated modulus and complex phase of the flavour-violating NSI coupling $\varepsilon_{\mu\tau}$. A value of $\delta_{\mathrm{CP}}=197^\circ$ from NuFit5.0 was fixed for all fits. The observed upper bound on the modulus results in $|\varepsilon_{\mu\tau}| \leq 5.4 \times 10^{-3}$ at 90\% CL, while no constraint on $\delta_{\mu\tau}$ can be placed at any CL. This result can be understood in the light of eqs.~\eqref{eq:NSImumu} and \eqref{eq:NSImue}, where $|\varepsilon_{\alpha\beta}|$ and $\delta_{\alpha\beta}$ appear directly coupled at first order for the three off-diagonal couplings, essentially rendering atmospheric neutrino experiments almost insensitive to the complex phases. However, the interplay of the two variables can impact the ability to constrain the modulus when the phase is profiled over, in comparison to the real-valued assumption.

%Observed likelihood ratio contour at 90\% CL of the complex, flavour-violating NSI coupling $\varepsilon_{\mu\tau}$. The shaded area indicates the allowed region of the parameter space $\{|\varepsilon_{\mu\tau}|,\delta_{\mu\tau}\}$ at the reported CL, containing the best fit point under the NSI hypothesis represented by a cross. The top and side panels show the projections of $-2\Delta \log \mathcal{L}$ when the other variable is profiled over. The dashed and solid grey lines of the adjoint projections are the 68\% and 90\% reference CL of the $\chi^2$ distribution with one degree of freedom.

\begin{figure}[h!]
\centering
\subfloat[$\varepsilon_{\mu\tau}$]{
\centering
\label{fig:profa}
\includegraphics[height=6cm]{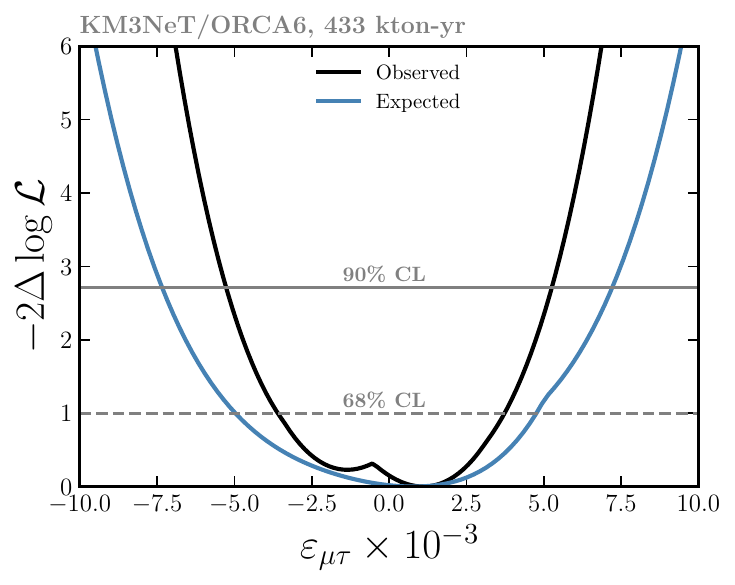}
}
\subfloat[$\varepsilon_{\tau\tau}-\varepsilon_{\mu\mu}$]{
\centering
\label{fig:profb}
\includegraphics[height=6cm]{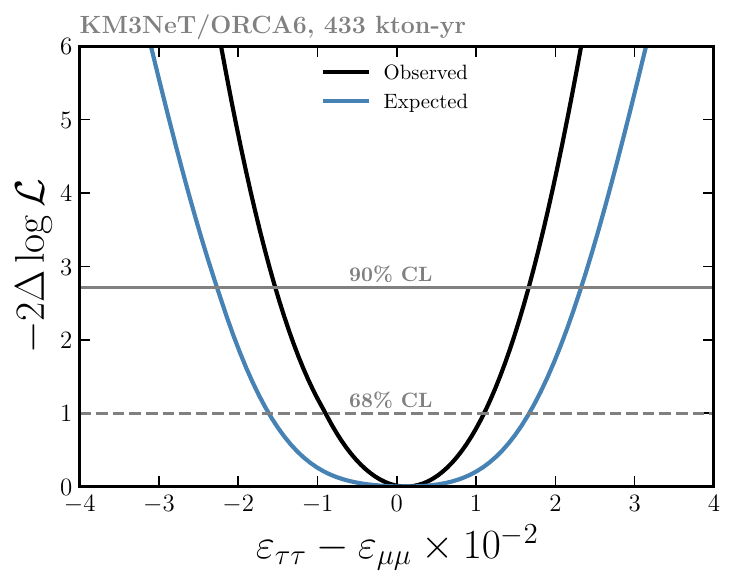}} 
% \subfloat[Profile in $\varepsilon_{e\tau}$]{
% \centering
% \label{fig:profc}
% \includegraphics[height=6cm]{imgs/results/profile_plot_Eps_et.pdf}}
% \subfloat[Profile in $\varepsilon_{e\mu}$]{
% \centering
% \label{fig:profd}
% \includegraphics[height=6cm]{imgs/results/profile_plot_Eps_em.pdf}}

\caption{ Observed log-likelihood ratio scan of the NSI parameters in the $\nu_\mu - \nu_\tau$ sector, where the flavour-violating coupling has been assumed real-valued by fixing its complex phase to 0 or $\pi$. Also shown are the expected sensitivities produced at the best fit from the SI case.}
\label{fig:profilesmutau}
\end{figure}

Figs. \ref{fig:profilesmutau} and \ref{fig:profileelec} present the one-dimensional $-2\Delta \log \mathcal{L}$ profiles of the NSI couplings observed from data (black) and expected when using MC generated at the best fit from SI (blue). In \ref{fig:profa}, the log-likelihood ratio is presented as a function of $\varepsilon_{\mu\tau}$ where the complex phase has been fixed to 0 or $\pi$ for comparison with other experiments. The observed allowed region at 90\% CL is found to be $[-5.3,5.3]\times 10^{-3}$. As can be seen in Fig.~\ref{fig:profa}, the profile exhibits a local minimum around $-2\times 10^{-3}$ corresponding to a local best-fit IO in opposition to the globally preferred NO. The inflection point in the log-likelihood ratio close to $\varepsilon_{\mu\tau}=0$ marks the transition from one to the opposite ordering. This feature is also seen in the profiles of Fig.~\ref{fig:profileelec} and is attributed to the degeneracies from eq.~\eqref{eq:vacuumtrans} and \eqref{eq:nsitrans}, which correlate the sign of $\Delta m^{2}_{31}$ to that of the real-valued NSI couplings or their complex phase.

%which is an inflection point in the behaviour of the likelihood. This point corresponds to an inversion of the best-fit NMO, attributed to the degeneracy explained in section \ref{sec:NSIprobs}, which causes the bounds on the parameter to be more restrictive in the semi-axis where the globally preferred NMO is located. This feature is common to all profiles of Figs. \ref{fig:profilesmutau} and \ref{fig:profileelec}.

%, while the FC corrections at that CL suggest that the interval could be up to 11\% more restrictive. Deviations from Wilks theorem are to be expected in beyond the Standard Model searches and neutrino oscillation analyses, as both parameters of interesest and nuisance parameters often run into effective boundaries of their phase space, and available statistics do not asymptotically reach the large sample limit  \cite{Algeri:2019lah}.

%\vspace{2cm}

Fig.~\ref{fig:profb} presents the observed log-likelihood profile corresponding to the non-universal coupling $\varepsilon_{\tau\tau}-\varepsilon_{\mu\mu}$. The sensitivity of ORCA6 is here an order of magnitude below that of $\varepsilon_{\mu\tau}$, as expected at probability level from the discussion in section \ref{sec:NSImutau}. The allowed confidence region is ${\varepsilon_{\tau\tau}-\varepsilon_{\mu\mu} \in [-1.5,1.7]\times 10^{-2}}$ at~90\%~CL. %, which could be up to 40\% more constraining after evaluation of the FC corrections shown in Fig.~\ref{fig:profb}. The PEs distribution suggests that the stronger FC corrections compared to the quantiles from Fig.~\ref{fig:profa} might stem from the proximity of the best-fit $\theta_{23}$ to maximal mixing. Under such circumstance, the muon survival probability being almost exactly reflected around $\varepsilon_{\tau\tau}-\varepsilon_{\mu\mu}=0$ renders such point an effective boundary of the parameter space, causing a stronger break-down of Wilks theorem assumptions.

\begin{figure}[h!]
\centering
\includegraphics[width=1.0\textwidth]{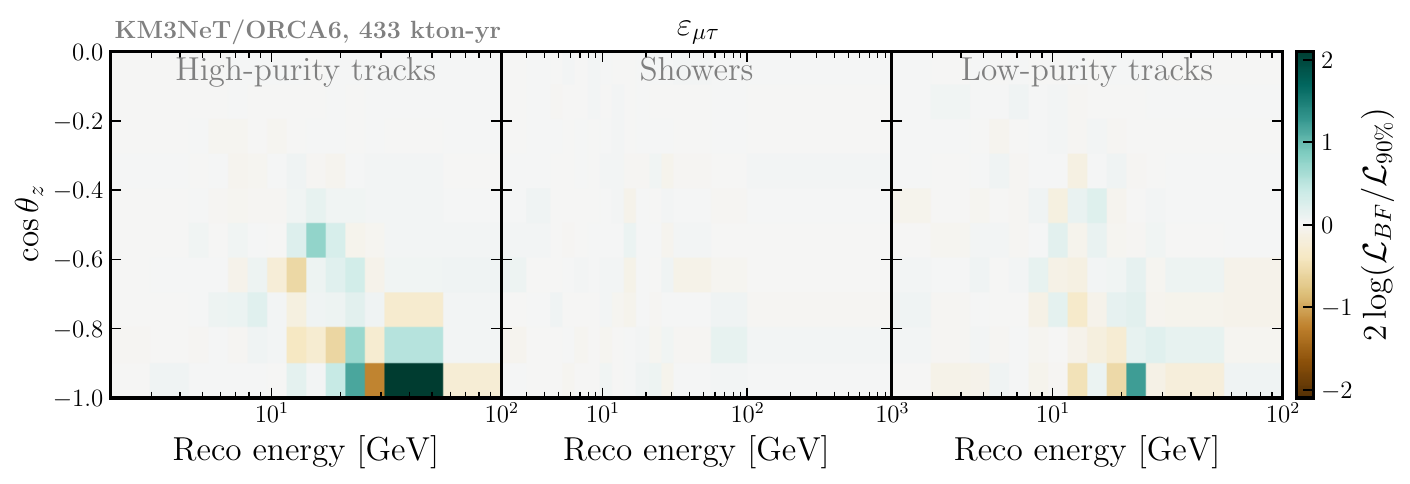}
\qquad
\caption{Bin-by-bin contribution to the log-likelihood ratio of the best-fit hypothesis over the point $|\varepsilon_{\mu\tau}|=0.0054$. From left to right, the three panels show the contributions arising from the high purity tracks, showers and low-purity tracks. As for Figs. \ref{fig:mapstautau}, \ref{fig:mapsetau} and \ref{fig:mapsemu}, positive values in the colour scale correspond to bins favouring the best fit over the chosen NSI coupling, thus contributing to the rejection power.}
\label{fig:mapsmutau}
\end{figure}

\begin{figure}[h!]
\centering
\includegraphics[width=1.0\textwidth]{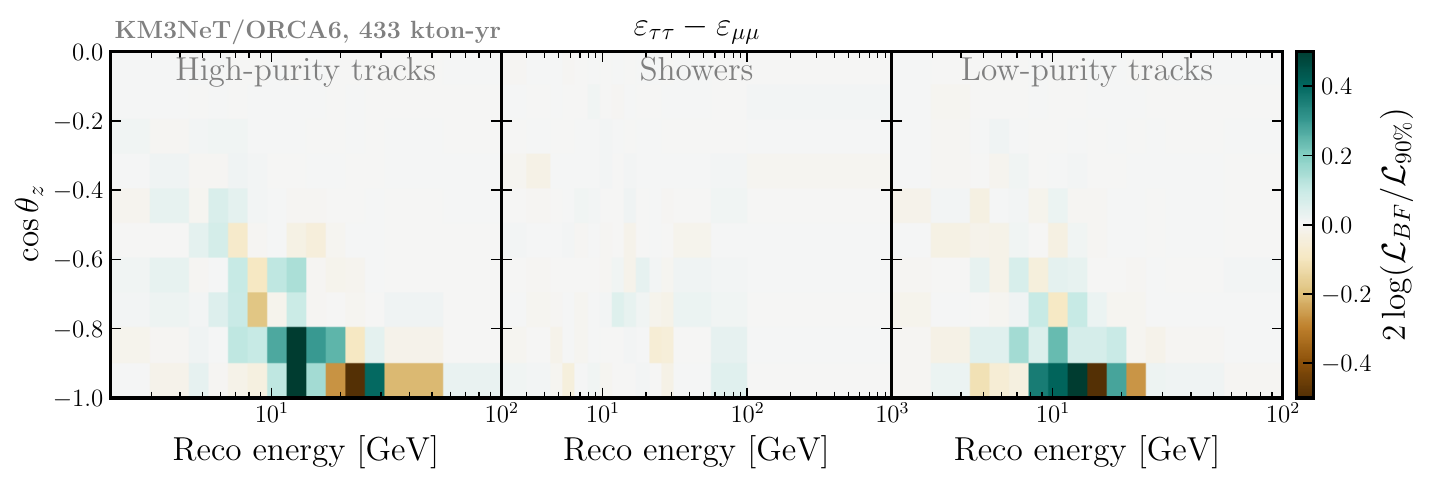}
\qquad
\caption{ Bin-by-bin contribution to the log-likelihood ratio of the best-fit hypothesis over the point $\varepsilon_{\tau\tau}-\varepsilon_{\mu\mu}=0.017$. From left to right, the three panels show the contributions arising from the high purity tracks, showers and low-purity tracks. }
\label{fig:mapstautau}
\end{figure}

The contribution of each region of reconstructed energy and direction to the constraints placed on the NSI couplings in the $\nu_{\mu}-\nu_{\tau}$ sector is shown in Figs.~\ref{fig:mapsmutau} and \ref{fig:mapstautau}. The colour scale represents the log-likelihood ratio of the best fit over an NSI point rejected at 90\% CL. Positive values correspond to bins favouring the best fit over the chosen NSI strength, consequently contributing to its rejection at 90\% CL. The most up-going neutrino baselines in the range $-1.0<\cos \theta_z < -0.8$ offer most of the sensitivity to both couplings, as strong matter effects present in core-crossing trajectories would enhance any NSI signature. The bin between [30, 55] GeV has the most prominent contribution to $\varepsilon_{\mu\tau}$-sensitivity. In fact, based on the MC distributions, this most-sensitive bin appears to be largely populated by neutrinos with true energies above 100 GeV.  The observation is consistent with Fig.~\ref{fig:NuMuOsc}, where the high-energy effects of $\varepsilon_{\mu\tau}$ are sustained across a wide range of energies above 100 GeV. The $L/E$ ratios seen in the left and central panels of Fig.~\ref{fig:LoE} reflect the same feature below 500 km/GeV; the high-energy event deficit driven by non-zero $\varepsilon_{\mu\tau}$ translates into $L/E$ ratios lower than the standard interaction case.

%arising from migrations of neutrinos in the range of 100$-$500 GeV of true energy which are reconstructed well below that due to limited detector resolution, and in fact dominate the population of that bin as suggested by the MC distribution

The log-likelihood ratio maps in Fig.~\ref{fig:mapstautau} correspond to a value of $\varepsilon_{\tau\tau}-\varepsilon_{\mu\mu}$ rejected at 90\%~CL over the best-fit hypothesis. The bin-by-bin contributions in both track sets follow the lines of constant $\nu_\mu /\bar{\nu}_\mu$-survival probability at the oscillation minimum, due to the oscillation damping caused by the non-zero NSI coupling in the model as seen in Fig.~\ref{fig:NuMuProbs}. Same effect is clear around $L/E\sim$1000 km/GeV in Fig.~\ref{fig:LoE} (left and centre).

% \begin{figure}[h!]
% \centering
% \includegraphics[width=1.0\textwidth]{imgs/results/merged_contours_paper-crop.pdf}
% \qquad
% \caption{Observed likelihood ratio contours at 90\% CL of the complex, flavour-violation NSI couplings. The shaded area indicates the allowed region of the parameter space $\{|\varepsilon_{\alpha\beta}|,\delta_{\alpha\beta}\}$ at the reported CL, containing the best fit point under each NSI hypothesis represented by a cross. The top and side panels show the projections of $-2\Delta \log \mathcal{L}$ when the other variable is profiled over.}
% \label{fig:complexmutau}
% \end{figure}

\newpage

\subsection{NSI in the $\nu_{e}$ sector}
\label{sec:elecsector}%f

\begin{figure}[h!]
\centering
\hspace{-0.5cm}
\subfloat[$\varepsilon_{e\tau}$]{
\centering
\label{fig:complexnsib}
\includegraphics[height=7.4cm]{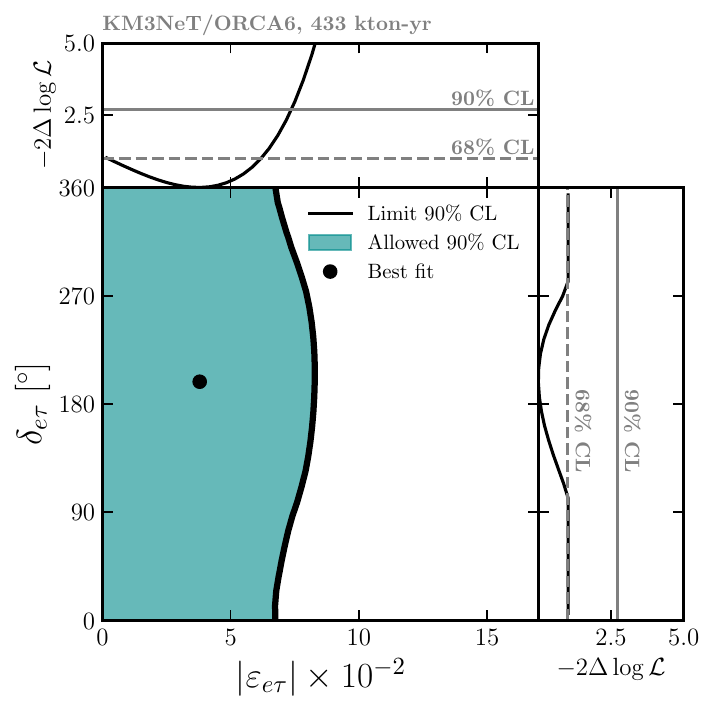}} 
\subfloat[$\varepsilon_{e\mu}$]{
\centering
\label{fig:complexnsic}
\includegraphics[height=7.4cm]{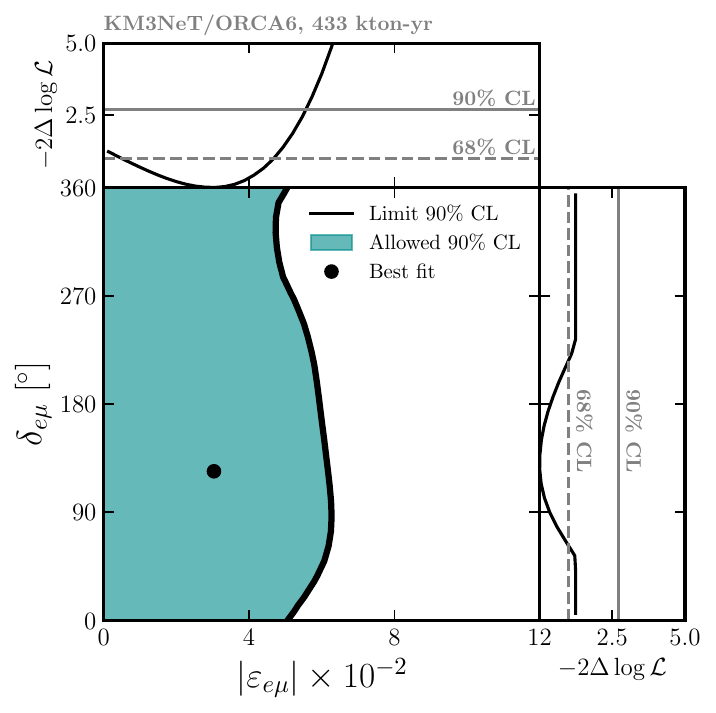}}

\caption{Observed log-likelihood ratio contours at 90\% CL of the complex, flavour-violating NSI couplings involving the electron flavour. The shaded area indicates the allowed region of the parameter space $\{|\varepsilon_{\alpha\beta}|,\delta_{\alpha\beta}\}$ at the reported CL, containing the best-fit point under each NSI hypothesis represented by a dot. The top and side panels show the projections of $-2\Delta \log \mathcal{L}$ when the other variable is profiled over.}
%\label{fig:complexnsia}
\label{fig:complexnsielec}
\end{figure}

\begin{figure}[h!]
\centering
% \subfloat[Profile in $\varepsilon_{\mu\tau}$]{
% \centering
% \label{fig:profa}
% \includegraphics[height=6cm]{imgs/results/profile_plot_Eps_mt.pdf}
% }
% \subfloat[Profile in $\varepsilon_{\tau\tau}-\varepsilon_{\mu\mu}$]{
% \centering
% \label{fig:profb}
% \includegraphics[height=6cm]{imgs/results/profile_plot_Eps_tt.pdf}} 
\subfloat[$\varepsilon_{e\tau}$]{
\centering
\label{fig:profc}
\includegraphics[height=6cm]{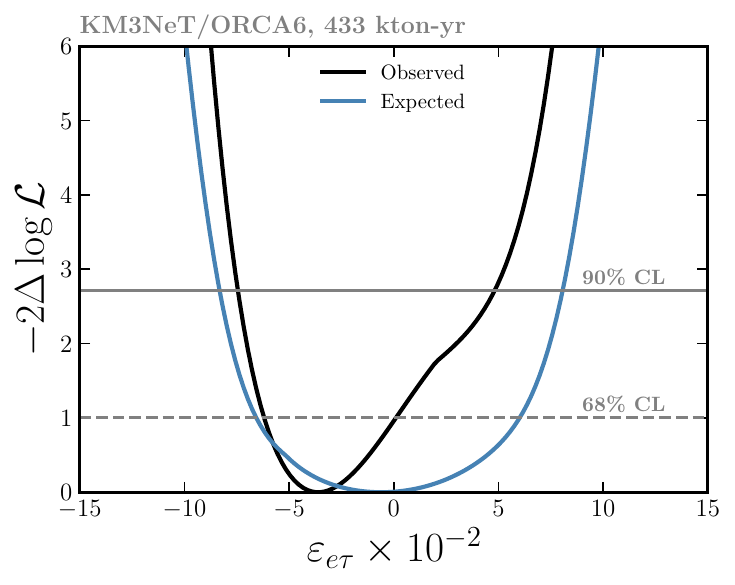}}
\subfloat[$\varepsilon_{e\mu}$]{
\centering
\label{fig:profd}
\includegraphics[height=6cm]{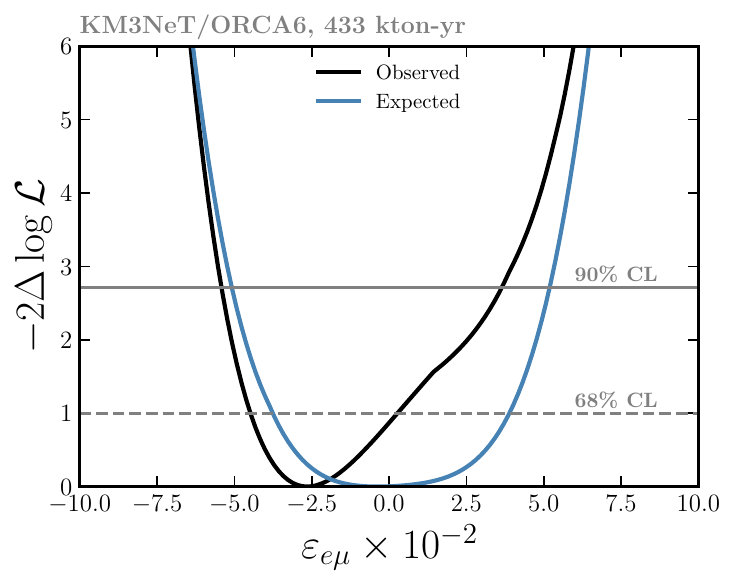}}

\caption{ Observed log-likelihood ratio profiles of the NSI parameters in the $\nu_e$ sector, where the couplings have been assumed real-valued by fixing their complex phase to 0 or $\pi$. Also shown are the expected sensitivities produced at the best fit from the SI case.}
\label{fig:profileelec}
\end{figure}

Fig.~\ref{fig:complexnsib} and \ref{fig:complexnsic} present the allowed regions for the modulus and complex phase of the NSI couplings involved at first order in $\nu_\mu \rightarrow \nu_e$ flavour transitions. The upper bounds are $|\varepsilon_{e\tau}| \leq 7.4 \times 10^{-2}$ and $|\varepsilon_{e\mu}| \leq 5.6 \times 10^{-2}$ at 90\% CL. As seen in Fig.~\ref{fig:profileelec} as well, in both cases the data was compatible with a best fit in the negative semi-axis, although the sensitivity does not reach to disfavour the positive one at 90\% CL. The best fit for $\varepsilon_{e\tau}$ and $\varepsilon_{e\mu}$ are consistent with a random fluctuation of the SM expectation. The complex-phase projections of the log-likelihood ratio shown in Figs. \ref{fig:complexnsib} and \ref{fig:complexnsic} reach a plateau at $-2\Delta \log \mathcal{L}=1$, since the NSI strength favoured by the data within this range reaches the boundary at zero. Both real-valued profiles of the parameters from Figs. \ref{fig:profc} and \ref{fig:profd} exhibit the degeneracy of eqs.~\eqref{eq:vacuumtrans} and \eqref{eq:nsitrans} that correlates their sign to that of the NMO, giving rise to the inflection points in the observed log-likelihood ratio profiles.

The extracted allowed regions under the real-valued assumption are $\varepsilon_{e\tau} \in [-7.4,4.8]\times 10^{-2}$ and $\varepsilon_{e\mu} \in [-5.4,3.6]\times 10^{-2}$, comparatively weaker than the bounds placed on the NSI couplings in the $\nu_\mu - \nu_\tau$ sector. 

\begin{figure}[h!]
\centering
\includegraphics[width=1.0\textwidth]{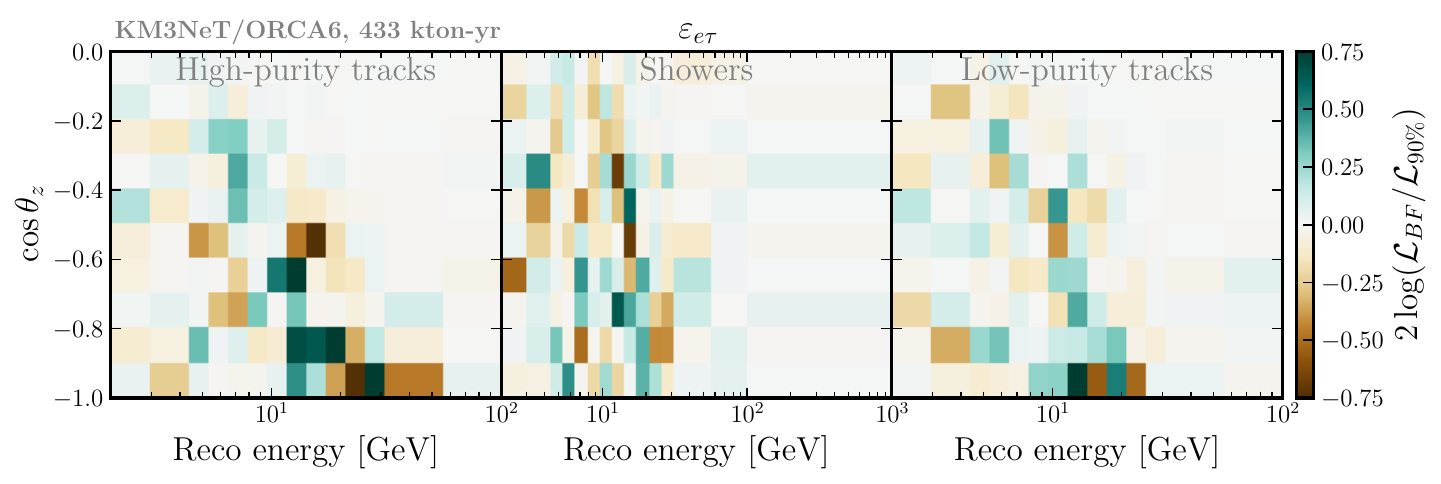}
\qquad
\caption{Bin-by-bin contribution to the log-likelihood ratio of the best-fit hypothesis over the point $|\varepsilon_{e\tau}|=0.075$. From left to right, the three panels show the contributions arising from the high-purity tracks, showers and low-purity tracks.  }
\label{fig:mapsetau}
\end{figure}

 In the log-likelihood ratio maps shown in Fig.~\ref{fig:mapsetau} the three event classes provide comparable contributions to the sensitivity to $\varepsilon_{e\tau}$, covering energy bins below 10 GeV consistent with the discussion in section \ref{sec:NSIelec}. NSI effects appear in the log-likelihood maps for a wider range of baselines compared to the couplings in the $\nu_\mu - \nu_\tau$ sector, driven by the periodic dependence on the matter potential not present in the latter case (see eq.~\eqref{eq:NSImue}). In the shower panel of the $L/E$ plot in Fig.~\ref{fig:LoE}, $\varepsilon_{e\tau}$ is the only coupling producing sizeable effects contributing to the 90\% CL sensitivity between 400 and 1000 km/GeV.

\begin{figure}[h!]
\centering
\includegraphics[width=1.0\textwidth]{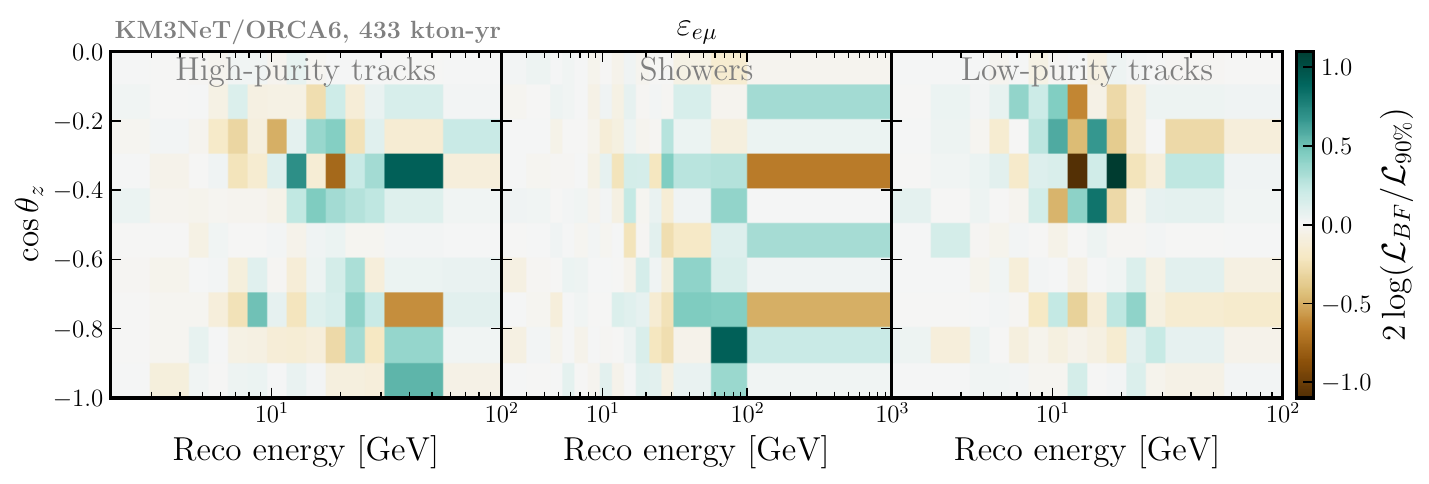}
\qquad
\caption{Bin-by-bin contribution to the log-likelihood ratio of the best-fit hypothesis over the point $|\varepsilon_{e\mu}|=0.055$. From left to right, the three panels show the contributions arising from the high-purity tracks, showers and low-purity tracks. }
\label{fig:mapsemu}
\end{figure}

The wide baseline coverage is common to the maps for an $\varepsilon_{e\mu}$ value rejected at 90\%~CL in Fig.~\ref{fig:mapsemu}. In this case, the three event samples start contributing at 10 GeV of reconstructed energy, in line with the effects shown in the $\varepsilon_{e\mu}$-panel of Figs.~\ref{fig:NuMuProbs} and \ref{fig:NuEProbs}. In contrast to $\varepsilon_{e\tau}$, the sensitivity to $\varepsilon_{e\mu}$ significantly benefits from high-energy effects appearing in the last energy bins above 30 GeV in the high-purity track and shower classes. 

\newpage

\subsection{Comparison of results}
\label{sec:comparison}

\begin{figure}[b!]
\centering
\includegraphics[width=0.6\textwidth]{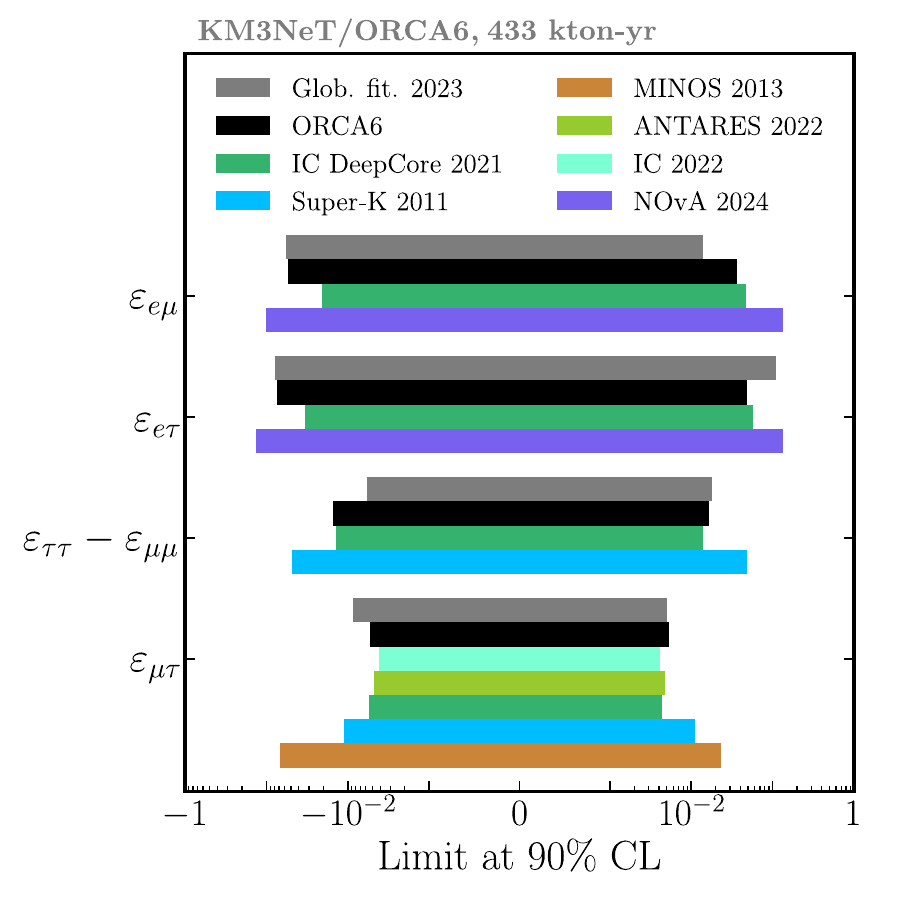}
\qquad
\caption{ 90\%CL constraints on the NSI couplings inferred from this study (ORCA6, 433 kton-years) in comparison to those reported by IC-DeepCore \cite{IceCubeCollaboration:2021euf},  MINOS \cite{MINOS:2013hmj}, NOvA \cite{NOvA:2024lti}, ANTARES \cite{ANTARES:2021crm}, IceCube (IC) \cite{IceCube:2022ubv} and Super-Kamiokande (Super-K) \cite{Super-Kamiokande:2011dam}, together with the global fit results from \cite{Coloma:2023ixt}. The latter accounts for correlations between NSI couplings by marginalising over all remaining parameters, and correlations are assumed in Super-K results in the $\mu-\tau$ sector as well. The bounds from IC-DeepCore 2021, MINOS 2013 and NOvA 2024 were re-scaled to down-quark NSI couplings.}
\label{fig:comparison}
\end{figure}

Fig.~\ref{fig:comparison} presents a summary of the reported allowed regions at 90\% CL of the NSI couplings, where the off-diagonal elements have been assumed real-valued with $\delta_{\alpha\beta}=0$ or $\pi$ for comparison. The results of IC-DeepCore 2021 \cite{IceCubeCollaboration:2021euf}, MINOS 2013 \cite{MINOS:2013hmj} and NOvA 2024 \cite{NOvA:2024lti} were re-scaled to match the NSI down-quark coupling convention used in this study (see eq.~\ref{eq:NSIfermion}). 

The ORCA6 limits on the four couplings investigated are of the same order and fully compatible with the current most stringent ones.  Large-volume neutrino telescopes benefit from a wide energy spectrum and baseline coverage to provide the leading limits to $\varepsilon_{\mu\tau}$ and $\varepsilon_{\tau\tau}-\varepsilon_{\mu\mu}$, and can additionally perform all-flavour analyses which strengthen the ability to constrain NSI in the $\nu_e$ sector. The strong matter effect experienced by neutrinos as they propagate along core-crossing trajectories enhances the sensitivity to deviations in the standard matter potential in neutrino telescopes, in comparison to long-baseline experiments like MINOS and NOvA. Among the former, ANTARES \cite{ANTARES:2021crm} and IceCube \cite{IceCube:2022ubv} have significantly higher energy thresholds than ORCA6. Therefore, they inspected $\varepsilon_{\mu\tau}$ only, due to its expected signatures in high-energy $\nu_\mu$ disappearance. Both report the allowed regions for NO and IO independently; the bar lengths in Fig.~\ref{fig:comparison} correspond to the boundary of the reported interval from the locally preferred mass ordering for the corresponding coupling sign, in alignment with the procedure of this work. In some cases, the bounds from ORCA6 are more stringent than the global fits to oscillation data from \cite{Coloma:2023ixt}, since the latter study assumes correlations among all NSI couplings which are not accounted for in the present work.

% ORCA6 synergy with osc minimum + high energy nu migration??
%%%%%%%%%%%%%%%%%%%%%%%%%%%%%%%%%%%%%%%%%%%%%%%%%%%%%%%%%%%%%%%%%%%%%%%%%%%%%%

%\input{conclusions.tex}
\section{Conclusions}
\label{sec:conclusions}
This work reports the results from the search for NC NSI with ORCA6, the configuration of the KM3NeT/ORCA neutrino telescope with the first six DUs. A total of 5828 events split into three classes and corresponding to an exposure of 433 kton-years are used. No significant deviation from standard interactions is found when measuring atmospheric neutrino oscillations in the GeV range as they propagate through the Earth. The observed data are found to be compatible at 90\% CL with the following bounds on the complex coupling strengths, inspected one-by-one assuming NSI in the neutrino forward scattering off down quarks:

%\newpage
% \begin{displaymath}
%     |\varepsilon_{\mu\tau}| \leq 5.4 \times 10^{-3}  \qquad \varepsilon_{\tau\tau}-\varepsilon_{\mu\mu} \in [-1.5,1.7]\times 10^{-2} 
% \end{displaymath}
% \begin{displaymath}
%     |\varepsilon_{e\tau}| \leq 7.4 \times 10^{-2} \qquad  |\varepsilon_{e\mu}| \leq 5.6 \times 10^{-2} \text ,
% \end{displaymath}

\begin{align*}
    |\varepsilon_{\mu\tau}| &\leq 5.4 \times 10^{-3}  & \varepsilon_{\tau\tau}-\varepsilon_{\mu\mu} &\in [-1.5,1.7]\times 10^{-2} \\
    |\varepsilon_{e\tau}| &\leq 7.4 \times 10^{-2}  & |\varepsilon_{e\mu}| &\leq 5.6 \times 10^{-2}
\end{align*}

No constraint can be placed on the complex phase of the flavour-violating couplings at 90\% CL. The above confidence regions align well with measurements provided by other neutrino experiments, and are of the same order as the current most stringent limits on the four inspected parameters. 

%resulting in stricter bounds compared to the Wilks theorem which is reported as the conservative choice. 

 An increased sensitivity to the individual NSI couplings was observed in certain regions of the phase space, as expected by theoretical predictions. The small statistical pulls on the nuisance parameters at the best fit indicate a proper understanding of the systematic uncertainties considered. Future analyses will conduct deeper studies on the detector response, atmospheric flux composition and neutrino interaction cross sections, in order to further reduce the relevant uncertainties. The current uncertainty in the measurement of the oscillation parameters obtained by ORCA6 has the highest impact on the ability to constrain NSI couplings, which will benefit from the measurement of $\theta_{23}$ and $\Delta m^{2}_{31}$ with improved detector resolution and increased statistics. 

These results constitute the first search for NSI conducted with KM3NeT. Future searches for NSI with an extended instrumented volume of ORCA will refine the statistical approach followed in this study,  including correlations among NSI couplings of all flavours therefore providing more general and model-independent constraints.

%BAYESIAN
%%%%%%%%%%%%%%%%%%%%%%%%%%%%%%%%%%%%%%%%%%%%%%%%%%%%%%%%%%%%%%%%%%%%%%%%%%%%%%%%%%%%%%%%%%%%%%%%%%%%%%%%%%%%%%%%%%%%%%%%%%%
%\input{acknowledgements.tex}
\section{Acknowledgements} 
The authors acknowledge the financial support of:
KM3NeT-INFRADEV2 project, funded by the European Union Horizon Europe Research and Innovation Programme under grant agreement No 101079679;
Funds for Scientific Research (FRS-FNRS), Francqui foundation, BAEF foundation;
Czech Science Foundation (GAČR 24-12702S);
Agence Nationale de la Recherche (contract ANR-15-CE31-0020), Centre National de la Recherche Scientifique (CNRS), Commission Europ\'eenne (FEDER fund and Marie Curie Program), LabEx UnivEarthS (ANR-10-LABX-0023 and ANR-18-IDEX-0001), Paris \^Ile-de-France Region, Normandy Region (Alpha, Blue-waves and Neptune), France,
For the CPER The Provence-Alpes-Côte d'Azur Delegation for Research and Innovation (DRARI), the Provence-Alpes-Côte d'Azur region, the Bouches-du-Rhône Departmental Council, the Metropolis of Aix-Marseille Provence and the City of Marseille through the CPER 2021-2027 NEUMED project,
The CNRS Institut National de Physique Nucléaire et de Physique des Particules (IN2P3); 
Shota Rustaveli National Science Foundation of Georgia (SRNSFG, FR-22-13708), Georgia;
This work is part of the MuSES project which has received funding from the European Research Council (ERC) under the European Union’s Horizon 2020 Research and Innovation Programme (grant agreement No 101142396).
The General Secretariat of Research and Innovation (GSRI), Greece;
Istituto Nazionale di Fisica Nucleare (INFN) and Ministero dell’Universit{\`a} e della Ricerca (MUR), through PRIN 2022 program (Grant PANTHEON 2022E2J4RK, Next Generation EU) and PON R\&I program (Avviso n. 424 del 28 febbraio 2018, Progetto PACK-PIR01 00021), Italy; IDMAR project Po-Fesr Sicilian Region az. 1.5.1; A. De Benedittis, W. Idrissi Ibnsalih, M. Bendahman, A. Nayerhoda, G. Papalashvili, I. C. Rea, A. Simonelli have been supported by the Italian Ministero dell'Universit{\`a} e della Ricerca (MUR), Progetto CIR01 00021 (Avviso n. 2595 del 24 dicembre 2019); KM3NeT4RR MUR Project National Recovery and Resilience Plan (NRRP), Mission 4 Component 2 Investment 3.1, Funded by the European Union – NextGenerationEU,CUP I57G21000040001, Concession Decree MUR No. n. Prot. 123 del 21/06/2022;
Ministry of Higher Education, Scientific Research and Innovation, Morocco, and the Arab Fund for Economic and Social Development, Kuwait;
Nederlandse organisatie voor Wetenschappelijk Onderzoek (NWO), the Netherlands;
Ministry of Research, Innovation and Digitalisation, Romania;
Slovak Research and Development Agency under Contract No. APVV-22-0413; Ministry of Education, Research, Development and Youth of the Slovak Republic;
MCIN for PID2021-124591NB-C41, -C42, -C43 and PDC2023-145913-I00 funded by MCIN/AEI/10.13039/501100011033 and by “ERDF A way of making Europe”, for ASFAE/2022/014 and ASFAE/2022 /023 with funding from the EU NextGenerationEU (PRTR-C17.I01) and Generalitat Valenciana, for Grant AST22\_6.2 with funding from Consejer\'{\i}a de Universidad, Investigaci\'on e Innovaci\'on and Gobierno de Espa\~na and European Union - NextGenerationEU, for CSIC-INFRA23013 and for CNS2023-144099, Generalitat Valenciana for CIDEGENT/2018/034, /2019/043, /2020/049, /2021/23, for CIDEIG/2023/20, for CIPROM/2023/51 and for GRISOLIAP/2021/192 and EU for MSC/101025085, Spain;
Khalifa University internal grants (ESIG-2023-008 and RIG-2023-070), United Arab Emirates;
The European Union's Horizon 2020 Research and Innovation Programme (ChETEC-INFRA - Project no. 101008324).
%%%%%%%%%%%%%%%%%%%%%%%%%%%%%%%%%%%%%%%%%%%%%%%%%%%%%%%%%%%%%%%%%%%%%%%%%%%%%%%%%%%%%%
%\newpage
%\input{appendix.tex}
\appendix

\section{Impact of nuisance parameters}
\label{app:nuisance}

\begin{figure}[h!]
\centering
\subfloat[$|\varepsilon_{\mu\tau}|$]{
\centering
\label{shifta}
\includegraphics[height=8.5cm]{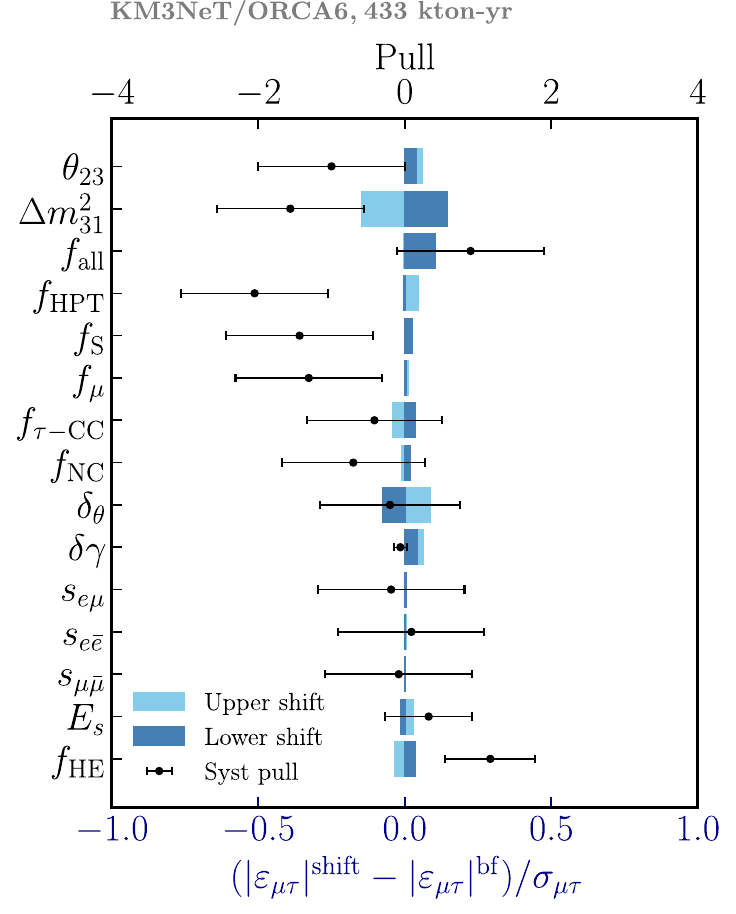} \
}
\subfloat[$\varepsilon_{\tau\tau}-\varepsilon_{\mu\mu}$]{
\centering
\label{shiftb}
\includegraphics[height=8.5cm]{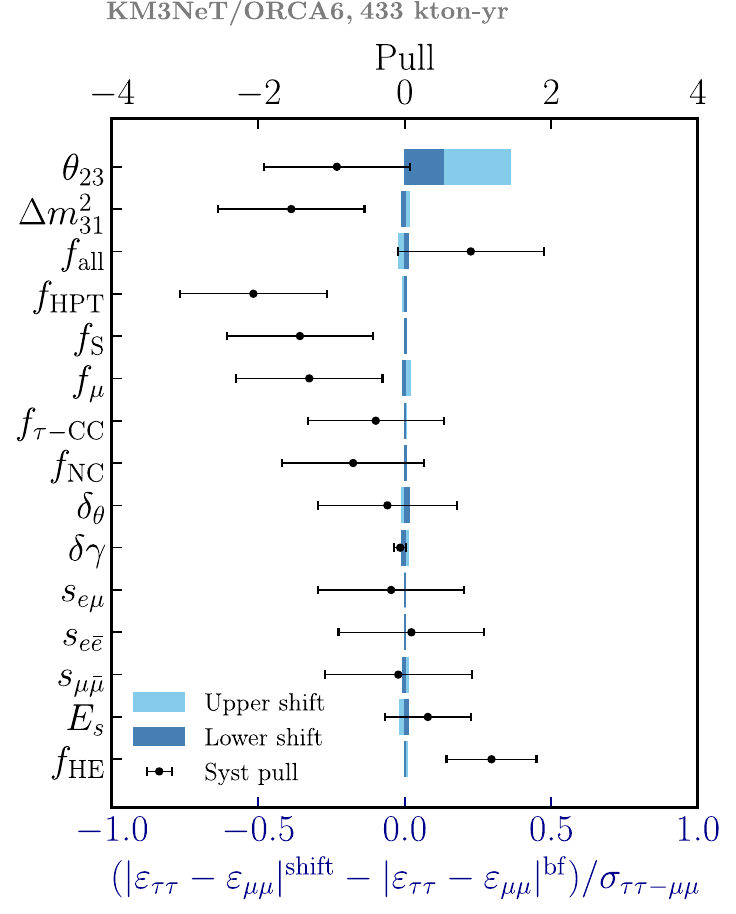}} 
% \subfloat[$\varepsilon_{e\tau}$]{
% \centering
% \label{shiftc}
% \includegraphics[height=7cm]{imgs/results/pulls_shift_data_Eps_et_grouped.pdf}}
% \subfloat[$\varepsilon_{e\mu}$]{
% \centering
% \label{shiftd}
% \includegraphics[height=7cm]{imgs/results/pulls_shift_data_Eps_em_grouped.pdf}}

\caption{ Shift of the systematics for the NSI coupling strengths $|\varepsilon_{\mu\tau}| $ and $\varepsilon_{\tau\tau}-\varepsilon_{\mu\mu}$. The blue bars reflect the shift induced on the PoI by the upper and lower $1\sigma$-shift of the systematic. The bars are to be read with the lower, blue axis. The overlaid black dots are the pulls experienced by the systematics in the NSI best fit, $\mathrm{(syst^{bf}-syst^{nom})/\sigma^{syst}}$, where $\mathrm{syst^{nom}}$ are the nominal values, and $\mathrm{\sigma^{syst}}$ is the prior width for constrained systematics, or the post-fit $1\sigma$ uncertainty for unconstrained ones. The pulls are to be read with the upper, black axis. The horizontal error bars on the pulls are the ratio of the post-fit uncertainty to the prior width. For unconstrained parameters which do not have priors, the error bars are set to one unit. The explanation of the systematic uncertainties is covered in Table \ref{tab:nuisance}.}
\label{fig:shifts1}
\end{figure}

\begin{figure}[h!]
\centering 
\subfloat[$|\varepsilon_{e\tau}|$]{
\centering
\label{shiftc}
\includegraphics[height=8.5cm]{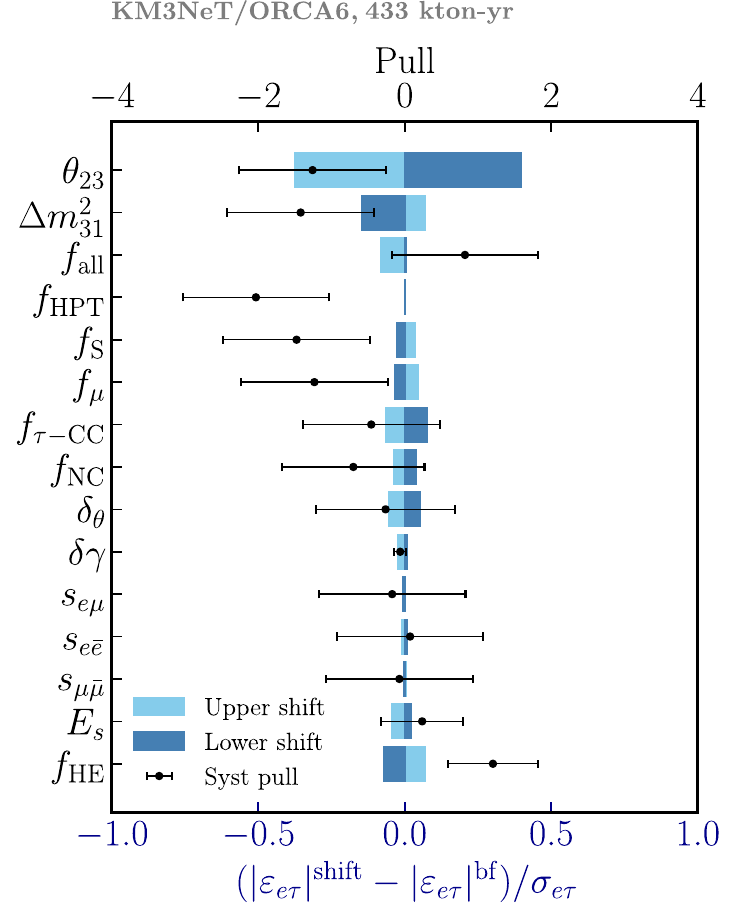}}
\subfloat[$|\varepsilon_{e\mu}|$]{
\centering
\label{shiftd}
\includegraphics[height=8.5cm]{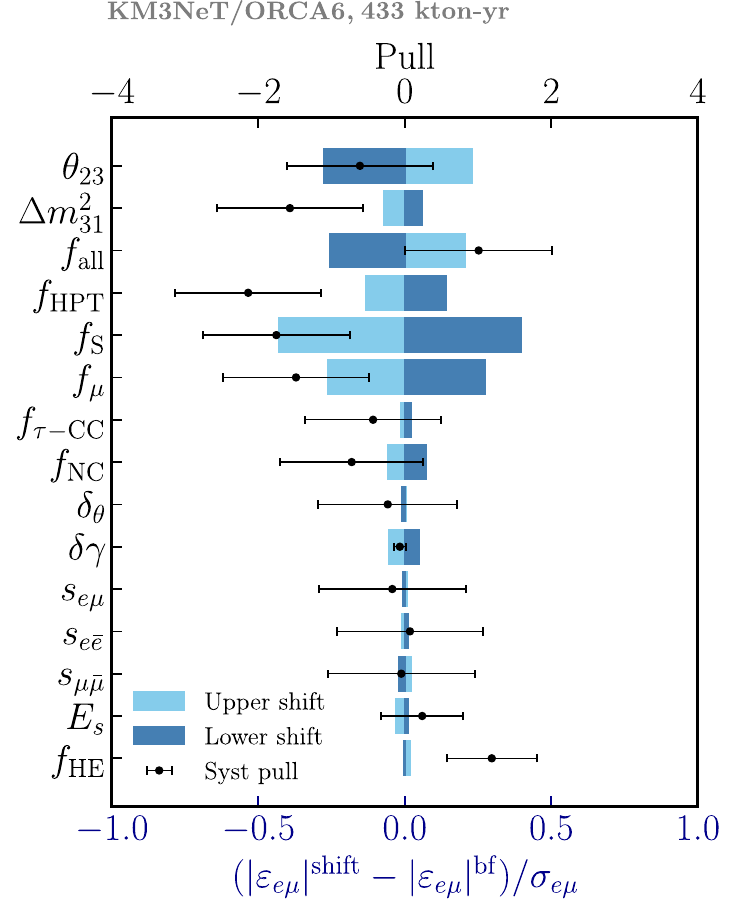}}

\caption{ Shift of the systematics for the NSI coupling strengths $|\varepsilon_{e\tau}| $ and $|\varepsilon_{e\mu}| $. The blue bars reflect the shift induced on the PoI by the upper and lower $1\sigma$-shift of the systematic. The bars are to be read with the lower, blue axis. The overlaid black dots are the pulls experienced by the systematics in the NSI best fit, $\mathrm{(syst^{bf}-syst^{nom})/\sigma^{syst}}$, where $\mathrm{syst^{nom}}$ are the nominal values, and $\mathrm{\sigma^{syst}}$ is the prior width for constrained systematics, or the post-fit $1\sigma$ uncertainty for unconstrained ones. The pulls are to be read with the upper, black axis. The horizontal error bars on the pulls are the ratio of the post-fit uncertainty to the prior width. For unconstrained parameters which do not have priors, the error bars are set to one unit. The explanation of the systematic uncertainties is covered in Table \ref{tab:nuisance}.}
\label{fig:shifts2}
\end{figure}

In the following, the impact of nuisance parameters on the individual NSI coupling strengths is described in detail. The approach consists of shifting the value of one systematic at a time, $\pm 1 \sigma$ away from its best-fit value under the given NSI coupling, which is considered the parameter of interest of the fit, and repeating the fit to the PoI and the remaining systematics fixing the shifted nuisance parameter. The impact of the systematic shift is shown as a deviation in the PoI from its best fit to the data, divided by the $1\sigma$ uncertainty obtained in the profile of the PoI. Only deviations in the modulus of the NSI parameters are considered, given the reduced observed sensitivity to the complex phases. This procedure yields the bar plots shown in Figs.~\ref{shifta}, \ref{shiftb}, \ref{shiftc} and \ref{shiftd}. On top of the bars, the black dots show the pulls exhibited by the systematics in the NSI fits, where all nuisance parameters are free to vary. For this, the central values of the oscillation parameters are those of NUFit 5.0 for NO \cite{Esteban:2020cvm},  since $\Delta m^{2}_{31} >0$ is preferred for the four NSI fits. As can be seen, all constrained nuisance parameters are found well within their expectation.

Figs.~\ref{fig:shifts1} and \ref{fig:shifts2} illustrate that the oscillation parameters $\Delta m^2_{31}$ and $\theta_{23}$ have the largest impact among all nuisance parameters, followed by the class normalisations, based on their shifts induced on the PoI. In particular, $\Delta m^2_{31}$ appears as the most important systematic for $\varepsilon_{\mu\tau}$, due to both parameters being able to shift the position of the oscillation valley. The mixing angle $\theta_{23}$ has the main impact for $\varepsilon_{\tau\tau}-\varepsilon_{\mu\mu}$ and $\varepsilon_{e\tau}$, driven by similar effects on the oscillation valley amplitude when varying the value of $\theta_{23}$ and $\varepsilon_{\tau\tau}-\varepsilon_{\mu\mu}$ or $\varepsilon_{e\tau}$. Finally, $\varepsilon_{e\mu}$ is most impacted by the shower class~normalisation.

%%%%%%%%%%%%%%%%%%%%%%%%%%%%%%%%%%%%%%%%%%%%%%%%%%%%%%%%%%%%%%%%%%%%%%%%%%

\clearpage
\bibliographystyle{JHEP}
\bibliography{main}

\providecommand{\href}[2]{#2}\begingroup\raggedright\begin{thebibliography}{10}

\bibitem{SNO:2001kpb}
{\scshape SNO} collaboration, \emph{{Measurement of the rate of $\nu_e+d \to
  p+p+e^-$ interactions produced by $^8$B solar neutrinos at the Sudbury
  Neutrino Observatory}},
  \href{https://doi.org/10.1103/PhysRevLett.87.071301}{\emph{Phys. Rev. Lett.}
  {\bfseries 87} (2001) 071301}
  [\href{https://arxiv.org/abs/nucl-ex/0106015}{{\ttfamily nucl-ex/0106015}}].

\bibitem{Super-Kamiokande:1998kpq}
{\scshape Super-Kamiokande} collaboration, \emph{{Evidence for oscillation of
  atmospheric neutrinos}},
  \href{https://doi.org/10.1103/PhysRevLett.81.1562}{\emph{Phys. Rev. Lett.}
  {\bfseries 81} (1998) 1562}
  [\href{https://arxiv.org/abs/hep-ex/9807003}{{\ttfamily hep-ex/9807003}}].

\bibitem{MACRO:1998ckv}
{\scshape MACRO} collaboration, \emph{{Measurement of the atmospheric neutrino
  induced upgoing muon flux using MACRO}},
  \href{https://doi.org/10.1016/S0370-2693(98)00885-5}{\emph{Phys. Lett. B}
  {\bfseries 434} (1998) 451}
  [\href{https://arxiv.org/abs/hep-ex/9807005}{{\ttfamily hep-ex/9807005}}].

\bibitem{Denton:2022een}
P.B.~Denton, M.~Friend, M.D.~Messier, H.A.~Tanaka, S.~B\"oser, J.a.A.B.~Coelho
  et~al., \emph{{Snowmass Neutrino Frontier: NF01 Topical Group Report on
  Three-Flavor Neutrino Oscillations}},
  \href{https://arxiv.org/abs/2212.00809}{{\ttfamily 2212.00809}}.

\bibitem{ParticleDataGroup:2024cfk}
{\scshape Particle Data Group} collaboration, \emph{{Review of particle
  physics}}, \href{https://doi.org/10.1103/PhysRevD.110.030001}{\emph{Phys.
  Rev. D} {\bfseries 110} (2024) 030001}.

\bibitem{Esteban:2020cvm}
I.~Esteban, M.C.~Gonzalez-Garcia, M.~Maltoni, T.~Schwetz and A.~Zhou,
  \emph{{The fate of hints: updated global analysis of three-flavor neutrino
  oscillations}}, \href{https://doi.org/10.1007/JHEP09(2020)178}{\emph{JHEP}
  {\bfseries 09} (2020) 178}
  [\href{https://arxiv.org/abs/2007.14792}{{\ttfamily 2007.14792}}].

\bibitem{deSalas:2020pgw}
P.F.~de~Salas, D.V.~Forero, S.~Gariazzo, P.~Mart\'\i{}nez-Mirav\'e, O.~Mena,
  C.A.~Ternes et~al., \emph{{2020 global reassessment of the neutrino
  oscillation picture}},
  \href{https://doi.org/10.1007/JHEP02(2021)071}{\emph{JHEP} {\bfseries 02}
  (2021) 071} [\href{https://arxiv.org/abs/2006.11237}{{\ttfamily
  2006.11237}}].

\bibitem{Capozzi:2021fjo}
F.~Capozzi, E.~Di~Valentino, E.~Lisi, A.~Marrone, A.~Melchiorri and A.~Palazzo,
  \emph{{Unfinished fabric of the three neutrino paradigm}},
  \href{https://doi.org/10.1103/PhysRevD.104.083031}{\emph{Phys. Rev. D}
  {\bfseries 104} (2021) 083031}
  [\href{https://arxiv.org/abs/2107.00532}{{\ttfamily 2107.00532}}].

\bibitem{Langacker:2010zza}
P.~Langacker, \emph{{The Standard Model and Beyond}}, Taylor \& Francis (2017),
  \href{https://doi.org/10.1201/b22175}{10.1201/b22175}.

\bibitem{Bischer:2019ttk}
I.~Bischer and W.~Rodejohann, \emph{{General neutrino interactions from an
  effective field theory perspective}},
  \href{https://doi.org/10.1016/j.nuclphysb.2019.114746}{\emph{Nucl. Phys. B}
  {\bfseries 947} (2019) 114746}
  [\href{https://arxiv.org/abs/1905.08699}{{\ttfamily 1905.08699}}].

\bibitem{Weinberg:1979sa}
S.~Weinberg, \emph{{Baryon and Lepton Nonconserving Processes}},
  \href{https://doi.org/10.1103/PhysRevLett.43.1566}{\emph{Phys. Rev. Lett.}
  {\bfseries 43} (1979) 1566}.

\bibitem{Grzadkowski:2010es}
B.~Grzadkowski, M.~Iskrzynski, M.~Misiak and J.~Rosiek, \emph{{Dimension-Six
  Terms in the Standard Model Lagrangian}},
  \href{https://doi.org/10.1007/JHEP10(2010)085}{\emph{JHEP} {\bfseries 10}
  (2010) 085} [\href{https://arxiv.org/abs/1008.4884}{{\ttfamily 1008.4884}}].

\bibitem{Farzan:2017xzy}
Y.~Farzan and M.~Tortola, \emph{{Neutrino oscillations and Non-Standard
  Interactions}}, \href{https://doi.org/10.3389/fphy.2018.00010}{\emph{Front.
  in Phys.} {\bfseries 6} (2018) 10}
  [\href{https://arxiv.org/abs/1710.09360}{{\ttfamily 1710.09360}}].

\bibitem{Chatterjee:2014gxa}
A.~Chatterjee, P.~Mehta, D.~Choudhury and R.~Gandhi, \emph{{Testing nonstandard
  neutrino matter interactions in atmospheric neutrino propagation}},
  \href{https://doi.org/10.1103/PhysRevD.93.093017}{\emph{Phys. Rev. D}
  {\bfseries 93} (2016) 093017}
  [\href{https://arxiv.org/abs/1409.8472}{{\ttfamily 1409.8472}}].

\bibitem{Dziewonski:1981xy}
A.M.~Dziewonski and D.L.~Anderson, \emph{{Preliminary reference Earth model}},
  \href{https://doi.org/10.1016/0031-9201(81)90046-7}{\emph{Phys. Earth Planet.
  Interiors} {\bfseries 25} (1981) 297}.

\bibitem{Gonzalez-Garcia:2013usa}
M.C.~Gonzalez-Garcia and M.~Maltoni, \emph{{Determination of matter potential
  from global analysis of neutrino oscillation data}},
  \href{https://doi.org/10.1007/JHEP09(2013)152}{\emph{JHEP} {\bfseries 09}
  (2013) 152} [\href{https://arxiv.org/abs/1307.3092}{{\ttfamily 1307.3092}}].

\bibitem{Wolfenstein:1977ue}
L.~Wolfenstein, \emph{{Neutrino Oscillations in Matter}},
  \href{https://doi.org/10.1103/PhysRevD.17.2369}{\emph{Phys. Rev. D}
  {\bfseries 17} (1978) 2369}.

\bibitem{RevModPhys.61.937}
T.K.~Kuo and J.~Pantaleone, \emph{Neutrino oscillations in matter},
  \href{https://doi.org/10.1103/RevModPhys.61.937}{\emph{Rev. Mod. Phys.}
  {\bfseries 61} (1989) 937}.

\bibitem{Wang:2018dwk}
T.~Wang and Y.-L.~Zhou, \emph{{Neutrino nonstandard interactions as a portal to
  test flavor symmetries}},
  \href{https://doi.org/10.1103/PhysRevD.99.035039}{\emph{Phys. Rev. D}
  {\bfseries 99} (2019) 035039}
  [\href{https://arxiv.org/abs/1801.05656}{{\ttfamily 1801.05656}}].

\bibitem{Coloma:2023ixt}
P.~Coloma, M.C.~Gonzalez-Garcia, M.~Maltoni, J.a.P.~Pinheiro and S.~Urrea,
  \emph{{Global constraints on non-standard neutrino interactions with quarks
  and electrons}}, \href{https://doi.org/10.1007/JHEP08(2023)032}{\emph{JHEP}
  {\bfseries 08} (2023) 032}
  [\href{https://arxiv.org/abs/2305.07698}{{\ttfamily 2305.07698}}].

\bibitem{Coloma:2016gei}
P.~Coloma and T.~Schwetz, \emph{{Generalized mass ordering degeneracy in
  neutrino oscillation experiments}},
  \href{https://doi.org/10.1103/PhysRevD.94.055005}{\emph{Phys. Rev. D}
  {\bfseries 94} (2016) 055005}
  [\href{https://arxiv.org/abs/1604.05772}{{\ttfamily 1604.05772}}].

\bibitem{Esteban:2018ppq}
I.~Esteban, M.C.~Gonzalez-Garcia, M.~Maltoni, I.~Martinez-Soler and J.~Salvado,
  \emph{{Updated constraints on non-standard interactions from global analysis
  of oscillation data}},
  \href{https://doi.org/10.1007/JHEP08(2018)180}{\emph{JHEP} {\bfseries 08}
  (2018) 180} [\href{https://arxiv.org/abs/1805.04530}{{\ttfamily
  1805.04530}}].

\bibitem{Proceedings:2019qno}
P.S.~Bhupal~Dev et~al., \emph{{Neutrino Non-Standard Interactions: A Status
  Report}},  \href{https://arxiv.org/abs/1907.00991}{{\ttfamily 1907.00991}}.

\bibitem{Friedland:2004ah}
A.~Friedland, C.~Lunardini and M.~Maltoni, \emph{{Atmospheric neutrinos as
  probes of neutrino-matter interactions}},
  \href{https://doi.org/10.1103/PhysRevD.70.111301}{\emph{Phys. Rev. D}
  {\bfseries 70} (2004) 111301}
  [\href{https://arxiv.org/abs/hep-ph/0408264}{{\ttfamily hep-ph/0408264}}].

\bibitem{Feng:2019mno}
W.-J.~Feng, J.~Tang, T.-C.~Wang and Y.-X.~Zhou, \emph{{Nonstandard interactions
  versus planet-scale neutrino oscillations}},
  \href{https://doi.org/10.1103/PhysRevD.100.115034}{\emph{Phys. Rev. D}
  {\bfseries 100} (2019) 115034}
  [\href{https://arxiv.org/abs/1909.12674}{{\ttfamily 1909.12674}}].

\bibitem{Esmaili:2013fva}
A.~Esmaili and A.Y.~Smirnov, \emph{{Probing Non-Standard Interaction of
  Neutrinos with IceCube and DeepCore}},
  \href{https://doi.org/10.1007/JHEP06(2013)026}{\emph{JHEP} {\bfseries 06}
  (2013) 026} [\href{https://arxiv.org/abs/1304.1042}{{\ttfamily 1304.1042}}].

\bibitem{Kikuchi:2008vq}
T.~Kikuchi, H.~Minakata and S.~Uchinami, \emph{{Perturbation Theory of Neutrino
  Oscillation with Nonstandard Neutrino Interactions}},
  \href{https://doi.org/10.1088/1126-6708/2009/03/114}{\emph{JHEP} {\bfseries
  03} (2009) 114} [\href{https://arxiv.org/abs/0809.3312}{{\ttfamily
  0809.3312}}].

\bibitem{Maki:1962mu}
Z.~Maki, M.~Nakagawa and S.~Sakata, \emph{{Remarks on the unified model of
  elementary particles}}, \href{https://doi.org/10.1143/PTP.28.870}{\emph{Prog.
  Theor. Phys.} {\bfseries 28} (1962) 870}.

\bibitem{joao_coelho_2023_8074017}
J.~Coelho, R.~Pestes, A.~Domi, N.~Chowdhury, S.~Bourret, L.~Maderer et~al.,
  \emph{joaoabcoelho/oscprob: v1.6.1},  June, 2023.
\newblock 10.5281/zenodo.8074017.

\bibitem{KM3Net:2016zxf}
{\scshape KM3NeT} collaboration, \emph{{Letter of intent for KM3NeT 2.0}},
  \href{https://doi.org/10.1088/0954-3899/43/8/084001}{\emph{J. Phys. G}
  {\bfseries 43} (2016) 084001}
  [\href{https://arxiv.org/abs/1601.07459}{{\ttfamily 1601.07459}}].

\bibitem{KM3NeT:2022pnv}
{\scshape KM3NeT} collaboration, \emph{{The KM3NeT multi-PMT optical module}},
  \href{https://doi.org/10.1088/1748-0221/17/07/P07038}{\emph{JINST} {\bfseries
  17} (2022) P07038} [\href{https://arxiv.org/abs/2203.10048}{{\ttfamily
  2203.10048}}].

\bibitem{adrianmartinez:in2p3-01220200}
{\scshape KM3NeT} collaboration, \emph{{The prototype detection unit of the
  KM3NeT detector}},
  \href{https://doi.org/10.1140/epjc/s10052-015-3868-9}{\emph{{Eur. Phys. J.
  C}} {\bfseries 76} (2016) 54}.

\bibitem{Real:2019rcv}
{\scshape KM3NeT} collaboration, \emph{Nanobeacon: A time calibration device
  for the km3net neutrino telescope},
  \href{https://doi.org/https://doi.org/10.1016/j.nima.2022.167132}{\emph{Nucl.
  Instrum. Methods Phys. Res.} {\bfseries 1040} (2022) 167132}.

\bibitem{KM3NeT:2024ecf}
{\scshape KM3NeT} collaboration, \emph{{Measurement of neutrino oscillation
  parameters with the first six detection units of KM3NeT/ORCA}},
  \href{https://doi.org/10.1007/JHEP10(2024)206}{\emph{JHEP} {\bfseries 10}
  (2024) 206} [\href{https://arxiv.org/abs/2408.07015}{{\ttfamily
  2408.07015}}].

\bibitem{KM3NeT:2021ozk}
{\scshape KM3NeT} collaboration, \emph{{Determining the neutrino mass ordering
  and oscillation parameters with KM3NeT/ORCA}},
  \href{https://doi.org/10.1140/epjc/s10052-021-09893-0}{\emph{Eur. Phys. J. C}
  {\bfseries 82} (2022) 26} [\href{https://arxiv.org/abs/2103.09885}{{\ttfamily
  2103.09885}}].

\bibitem{KM3NeT:2021uez}
{\scshape KM3NeT} collaboration, \emph{{Sensitivity to light sterile neutrino
  mixing parameters with KM3NeT/ORCA}},
  \href{https://doi.org/10.1007/JHEP10(2021)180}{\emph{JHEP} {\bfseries 10}
  (2021) 180} [\href{https://arxiv.org/abs/2107.00344}{{\ttfamily
  2107.00344}}].

\bibitem{KM3NeT:2023ncz}
{\scshape KM3NeT} collaboration, \emph{{Probing invisible neutrino decay with
  KM3NeT/ORCA}}, \href{https://doi.org/10.1007/JHEP04(2023)090}{\emph{JHEP}
  {\bfseries 04} (2023) 090}
  [\href{https://arxiv.org/abs/2302.02717}{{\ttfamily 2302.02717}}].

\bibitem{BAKER1984437}
S.~Baker and R.D.~Cousins, \emph{Clarification of the use of chi-square and
  likelihood functions in fits to histograms},
  \href{https://doi.org/https://doi.org/10.1016/0167-5087(84)90016-4}{\emph{Nucl.
  Instrum. Methods Phys. Res.} {\bfseries 221} (1984) 437}.

\bibitem{Conway:2011in}
J.S.~Conway, \emph{{Incorporating Nuisance Parameters in Likelihoods for
  Multisource Spectra}},  in \emph{{PHYSTAT 2011}}, pp.~115--120, 2011,
  \href{https://doi.org/10.5170/CERN-2011-006.115}{DOI}
  [\href{https://arxiv.org/abs/1103.0354}{{\ttfamily 1103.0354}}].

\bibitem{Barlow:1993dm}
R.J.~Barlow and C.~Beeston, \emph{{Fitting using finite Monte Carlo samples}},
  \href{https://doi.org/10.1016/0010-4655(93)90005-W}{\emph{Comput. Phys.
  Commun.} {\bfseries 77} (1993) 219}.

\bibitem{Wilks:1938dza}
S.S.~Wilks, \emph{{The Large-Sample Distribution of the Likelihood Ratio for
  Testing Composite Hypotheses}},
  \href{https://doi.org/10.1214/aoms/1177732360}{\emph{Annals Math. Statist.}
  {\bfseries 9} (1938) 60}.

\bibitem{Algeri:2019lah}
S.~Algeri, J.~Aalbers, K.~Dundas~Mor\r{a} and J.~Conrad, \emph{{Searching for
  new phenomena with profile likelihood ratio tests}},
  \href{https://doi.org/10.1038/s42254-020-0169-5}{\emph{Nature Rev. Phys.}
  {\bfseries 2} (2020) 245} [\href{https://arxiv.org/abs/1911.10237}{{\ttfamily
  1911.10237}}].

\bibitem{Honda:2006qj}
M.~Honda, T.~Kajita, K.~Kasahara, S.~Midorikawa and T.~Sanuki,
  \emph{{Calculation of atmospheric neutrino flux using the interaction model
  calibrated with atmospheric muon data}},
  \href{https://doi.org/10.1103/PhysRevD.75.043006}{\emph{Phys. Rev. D}
  {\bfseries 75} (2007) 043006}
  [\href{https://arxiv.org/abs/astro-ph/0611418}{{\ttfamily
  astro-ph/0611418}}].

\bibitem{Barr:2006it}
G.D.~Barr, T.K.~Gaisser, S.~Robbins and T.~Stanev, \emph{{Uncertainties in
  Atmospheric Neutrino Fluxes}},
  \href{https://doi.org/10.1103/PhysRevD.74.094009}{\emph{Phys. Rev. D}
  {\bfseries 74} (2006) 094009}
  [\href{https://arxiv.org/abs/astro-ph/0611266}{{\ttfamily
  astro-ph/0611266}}].

\bibitem{PhysRevD.92.023004}
M.~Honda, M.S.~Athar, T.~Kajita, K.~Kasahara and S.~Midorikawa,
  \emph{Atmospheric neutrino flux calculation using the nrlmsise-00 atmospheric
  model}, \href{https://doi.org/10.1103/PhysRevD.92.023004}{\emph{Phys. Rev. D}
  {\bfseries 92} (2015) 023004}.

\bibitem{IceCubeCollaboration:2021euf}
{\scshape IceCube} collaboration, \emph{{All-flavor constraints on nonstandard
  neutrino interactions and generalized matter potential with three years of
  IceCube DeepCore data}},
  \href{https://doi.org/10.1103/PhysRevD.104.072006}{\emph{Phys. Rev. D}
  {\bfseries 104} (2021) 072006}
  [\href{https://arxiv.org/abs/2106.07755}{{\ttfamily 2106.07755}}].

\bibitem{MINOS:2013hmj}
{\scshape MINOS} collaboration, \emph{{Search for Flavor-Changing Non-Standard
  Neutrino Interactions by MINOS}},
  \href{https://doi.org/10.1103/PhysRevD.88.072011}{\emph{Phys. Rev. D}
  {\bfseries 88} (2013) 072011}
  [\href{https://arxiv.org/abs/1303.5314}{{\ttfamily 1303.5314}}].

\bibitem{NOvA:2024lti}
{\scshape NOvA} collaboration, \emph{{Search for CP-Violating Neutrino
  Nonstandard Interactions with the NOvA Experiment}},
  \href{https://doi.org/10.1103/PhysRevLett.133.201802}{\emph{Phys. Rev. Lett.}
  {\bfseries 133} (2024) 201802}
  [\href{https://arxiv.org/abs/2403.07266}{{\ttfamily 2403.07266}}].

\bibitem{ANTARES:2021crm}
{\scshape ANTARES} collaboration, \emph{{Search for non-standard neutrino
  interactions with 10 years of ANTARES data}},
  \href{https://doi.org/10.1007/JHEP07(2022)048}{\emph{JHEP} {\bfseries 07}
  (2022) 048} [\href{https://arxiv.org/abs/2112.14517}{{\ttfamily
  2112.14517}}].

\bibitem{IceCube:2022ubv}
{\scshape IceCube} collaboration, \emph{{Strong Constraints on Neutrino
  Nonstandard Interactions from TeV-Scale $\nu_u$ Disappearance at IceCube}},
  \href{https://doi.org/10.1103/PhysRevLett.129.011804}{\emph{Phys. Rev. Lett.}
  {\bfseries 129} (2022) 011804}
  [\href{https://arxiv.org/abs/2201.03566}{{\ttfamily 2201.03566}}].

\bibitem{Super-Kamiokande:2011dam}
{\scshape Super-Kamiokande} collaboration, \emph{{Study of Non-Standard
  Neutrino Interactions with Atmospheric Neutrino Data in Super-Kamiokande I
  and II}}, \href{https://doi.org/10.1103/PhysRevD.84.113008}{\emph{Phys. Rev.
  D} {\bfseries 84} (2011) 113008}
  [\href{https://arxiv.org/abs/1109.1889}{{\ttfamily 1109.1889}}].

\end{thebibliography}\endgroup
%\input{authors}

% \bibitem{Davis}
% Davis R Jr, Harmer DS, Hoffman KC.
% \emph{Search for neutrinos from the sun},
% \emph{Phys Rev Lett} {\bf 20} (1968) 1205–9.

% \bibitem{Kajita}
% Fukuda Y, et al.
% \emph{Evidence for oscillation of atmospheric neutrinos},
% \emph{Phys Rev Lett} {\bf 81} (1998) 1562–67.

% \bibitem{Snowmass}
% Denton, P. et. al.
% \emph{Snowmass Neutrino Frontier: NF01 Topical Group Report on Three-Flavor Neutrino Oscillations},
% arXiv:2212.00809.

% \bibitem{Nufit}
% Esteban, I., Gonzalez-Garcia, M., Maltoni, M. et al.
% \emph{ The fate of hints: updated global analysis of three-flavor neutrino oscillations},
% \emph{JHEP} {\bf 09} (2020) 178.

% \bibitem{GlobalValencia}
% de Salas, P.F., Forero, D.V., Gariazzo, S. et al.
% \emph{2020 global reassessment of the neutrino oscillation picture},
% \emph{JHEP} {\bf 02} (2021) 71.

%\end{thebibliography}
\end{document}